\pdfoutput=1
\documentclass[12pt,a4paper]{article}

\usepackage{siunitx, collcell}
\usepackage{ifthen} 
\newboolean{pdflatex}
\setboolean{pdflatex}{true} 

\newboolean{articletitles}
\setboolean{articletitles}{true} 

\newboolean{uprightparticles}
\setboolean{uprightparticles}{false} 
\usepackage{placeins}

\def\paperauthors{LHCb collaboration} 
\def\paperasciititle{Measurement of the CKM angle gamma with B->D(K3pi)h decays. } 
\def\papertitle{Measurement of the CKM angle $\gamma$ with $ B^\pm \to D[K^\mp \pi^\pm \pi^\pm \pi^\mp] h^\pm$ decays using a binned phase-space approach} 
\def\paperkeywords{{High Energy Physics}, {LHCb}} 
\def\papercopyright{\the\year\ CERN for the benefit of the LHCb collaboration} 
\def\paperlicence{CC BY 4.0 licence}
\def\paperlicenceurl{https://creativecommons.org/licenses/by/4.0/}

\def\KS      {\ensuremath{K^0_{\rm S}}}

\def\CP      {\ensuremath{\mathcal{CP}}}

\def\Cf      {\ensuremath{R_{K3\pi}}}
\def\sp      {\ensuremath{\delta_{K3\pi}}}
\def\cfi     {\ensuremath{R^i_{K3\pi}}}
\def\spi     {\ensuremath{\delta^i_{K3\pi}}}
\def\ri      {\ensuremath{r^i_{K3\pi}}} 
\newcommand{\aerr}[2]{{\:}^{+{\:}#1}_{-{\:}#2}}

\def\gammav  {\left( 54.8 \aerr{6.0}{5.8}\aerr{0.6}{0.6}\aerr{6.7}{4.3} \right)^\circ}

\def\xkv     {\left( -66.4 \aerr{7.6}{7.2} \aerr{0.9}{0.9}\aerr{10.0}{\phantom{0}7.3} \right)\times 10^{-3}}
\def\ykv     {\left( \phantom{-}67.4 \aerr{7.1}{7.5} \aerr{0.8}{1.0}\aerr{11.1}{12.1}\right)\times 10^{-3}}
\def\xpiv    {\left( \phantom{-1}3.0 \aerr{1.2}{1.2}\aerr{0.1}{0.1} \aerr{\phantom{0}0.9}{\phantom{0}0.9} \right)\times 10^{-3}}
\def\ypiv     {\left( -\phantom{1}3.4 \aerr{1.1}{1.1}\aerr{0.3}{0.3} \aerr{\phantom{0}0.8}{\phantom{0}0.8} \right)\times 10^{-3}}
\def\dBKv     {\left( 134.6 \aerr{\phantom{1}6.0}{\phantom{1}6.0}\aerr{0.7}{0.7}\aerr{\phantom{1}8.6}{\phantom{1}8.7} \right)^\circ}
\def\dBpv     {\left(  311.8 \aerr{14.7}{15.0}\aerr{3.0}{2.3}\aerr{14.7}{15.0} \right)^\circ }
\def\rBKv     {\left( \phantom{1}94.6 \aerr{\phantom{1}3.1}{\phantom{1}3.1}\aerr{0.5}{0.5}\aerr{\phantom{1}3.0}{\phantom{1}2.3}\right)\times10^{-3}}
\def\rBpv     {\left( \phantom{19}4.5\aerr{\phantom{1}1.1}{\phantom{1}1.0}\aerr{0.3}{0.3}\aerr{\phantom{1}0.4}{\phantom{1}0.3} \right)\times 10^{-3}}

\usepackage[top=1in, bottom=1.25in, left=1in, right=1in]{geometry}

\usepackage{microtype}
\usepackage{lineno}  
\usepackage{xspace} 
\usepackage{caption} 

\usepackage{graphicx}  
\usepackage{color}
\usepackage{colortbl}
\graphicspath{{./figs/}} 

\usepackage{amsmath} 
\usepackage{amssymb}
\usepackage{amsfonts}
\usepackage{upgreek} 

\newcommand*\patchAmsMathEnvironmentForLineno[1]{%
\expandafter\let\csname old#1\expandafter\endcsname\csname #1\endcsname
\expandafter\let\csname oldend#1\expandafter\endcsname\csname
end#1\endcsname
 \renewenvironment{#1}%
   {\linenomath\csname old#1\endcsname}%
   {\csname oldend#1\endcsname\endlinenomath}%
}
\newcommand*\patchBothAmsMathEnvironmentsForLineno[1]{%
  \patchAmsMathEnvironmentForLineno{#1}%
  \patchAmsMathEnvironmentForLineno{#1*}%
}
\AtBeginDocument{%
\patchBothAmsMathEnvironmentsForLineno{equation}%
\patchBothAmsMathEnvironmentsForLineno{align}%
\patchBothAmsMathEnvironmentsForLineno{flalign}%
\patchBothAmsMathEnvironmentsForLineno{alignat}%
\patchBothAmsMathEnvironmentsForLineno{gather}%
\patchBothAmsMathEnvironmentsForLineno{multline}%
\patchBothAmsMathEnvironmentsForLineno{eqnarray}%
}

\usepackage{hyperxmp}

\usepackage[pdftex,
            pdfauthor={\paperauthors},
            pdftitle={\paperasciititle},
            pdfkeywords={\paperkeywords},
            pdfcopyright={Copyright (C) \papercopyright},
            pdflicenseurl={\paperlicenceurl}]{hyperref}

\usepackage[colorinlistoftodos,textsize=scriptsize]{todonotes}

\usepackage[bottom,flushmargin,hang,multiple]{footmisc}

\usepackage[all]{hypcap} 

\usepackage{xspace} 
\usepackage{upgreek}

\def\lhcb   {\mbox{LHCb}\xspace}

\def\MagUp {\mbox{\em Mag\kern -0.05em Up}\xspace}

\ifthenelse{\boolean{uprightparticles}}%
{

 \def\Pnu         {\ensuremath{\upnu}\xspace}                 
                  
 \def\Ppi         {\ensuremath{\uppi}\xspace}

 \def\PDelta      {\ensuremath{\Delta}\xspace}                 
 \def\PXi         {\ensuremath{\Xi}\xspace}                 
 \def\PLambda     {\ensuremath{\Lambda}\xspace}                 
 \def\PSigma      {\ensuremath{\Sigma}\xspace}                 
 \def\POmega      {\ensuremath{\Omega}\xspace}                 
 \def\PUpsilon    {\ensuremath{\Upsilon}\xspace}

 \def\PB      {\ensuremath{\mathrm{B}}\xspace}                 
                  
 \def\PD      {\ensuremath{\mathrm{D}}\xspace}

 \def\PK      {\ensuremath{\mathrm{K}}\xspace}

 \def\Pb      {\ensuremath{\mathrm{b}}\xspace}                 
 \def\Pc      {\ensuremath{\mathrm{c}}\xspace}                 
 \def\Pd      {\ensuremath{\mathrm{d}}\xspace}

 \def\Pi      {\ensuremath{\mathrm{i}}\xspace}

 \def\Ps      {\ensuremath{\mathrm{s}}\xspace}                 
                  
 \def\Pu      {\ensuremath{\mathrm{u}}\xspace}

 \def\thebaroffset{0.0em}
}
{

 \def\Pnu         {\ensuremath{\nu}\xspace}                 
                  
 \def\Ppi         {\ensuremath{\pi}\xspace}

 \mathchardef\PDelta="7101
 \mathchardef\PXi="7104
 \mathchardef\PLambda="7103
 \mathchardef\PSigma="7106
 \mathchardef\POmega="710A
 \mathchardef\PUpsilon="7107
                  
 \def\PB      {\ensuremath{B}\xspace}                 
                  
 \def\PD      {\ensuremath{D}\xspace}

 \def\PK      {\ensuremath{K}\xspace}

 \def\Pb      {\ensuremath{b}\xspace}                 
 \def\Pc      {\ensuremath{c}\xspace}                 
 \def\Pd      {\ensuremath{d}\xspace}

 \def\Pi      {\ensuremath{i}\xspace}

 \def\Ps      {\ensuremath{s}\xspace}                 
                  
 \def\Pu      {\ensuremath{u}\xspace}

 \def\thebaroffset{0.18em}
}
\newcommand{\offsetoverline}[2][\thebaroffset]{\kern #1\overline{\kern -#1 #2}}%

\makeatletter
\ifcase \@ptsize \relax
  \newcommand{\miniscule}{\@setfontsize\miniscule{4}{5}}
\or
  \newcommand{\miniscule}{\@setfontsize\miniscule{5}{6}}
\or
  \newcommand{\miniscule}{\@setfontsize\miniscule{5}{6}}
\fi
\makeatother

\DeclareRobustCommand{\optbar}[1]{\shortstack{{\miniscule (\rule[.5ex]{1.25em}{.18mm})}
  \\ [-.7ex] $#1$}}

\def\neub       {{\ensuremath{\overline{\Pnu}}}\xspace}

\def\neumb      {{\ensuremath{\neub_\mu}}\xspace}

\def\uquark    {{\ensuremath{\Pu}}\xspace}

\def\dquark    {{\ensuremath{\Pd}}\xspace}

\def\squark    {{\ensuremath{\Ps}}\xspace}

\def\cquark    {{\ensuremath{\Pc}}\xspace}

\def\bquark    {{\ensuremath{\Pb}}\xspace}

\def\pion   {{\ensuremath{\Ppi}}\xspace}
\def\piz    {{\ensuremath{\pion^0}}\xspace}
\def\pip    {{\ensuremath{\pion^+}}\xspace}
\def\pim    {{\ensuremath{\pion^-}}\xspace}
\def\pipm   {{\ensuremath{\pion^\pm}}\xspace}
\def\pimp   {{\ensuremath{\pion^\mp}}\xspace}

\def\kaon    {{\ensuremath{\PK}}\xspace}

\def\KorKbar {\kern \thebaroffset\optbar{\kern -\thebaroffset \PK}{}\xspace}

\def\Kp      {{\ensuremath{\kaon^+}}\xspace}
\def\Km      {{\ensuremath{\kaon^-}}\xspace}
\def\Kpm     {{\ensuremath{\kaon^\pm}}\xspace}
\def\Kmp     {{\ensuremath{\kaon^\mp}}\xspace}
\def\KS      {{\ensuremath{\kaon^0_{\mathrm{S}}}}\xspace}

\def\Dbar    {{\ensuremath{\offsetoverline{\PD}}}\xspace}
\def\D       {{\ensuremath{\PD}}\xspace}

\def\DorDbar {\kern \thebaroffset\optbar{\kern -\thebaroffset \PD}\xspace}
\def\Dz      {{\ensuremath{\D^0}}\xspace}
\def\Dzb     {{\ensuremath{\Dbar{}^0}}\xspace}
\def\Dp      {{\ensuremath{\D^+}}\xspace}
\def\Dm      {{\ensuremath{\D^-}}\xspace}

\def\DpDm    {\ensuremath{\Dp {\kern -0.16em \Dm}}\xspace}

\def\B       {{\ensuremath{\PB}}\xspace}

\def\BorBbar {\kern \thebaroffset\optbar{\kern -\thebaroffset \PB}\xspace}
\def\Bz      {{\ensuremath{\B^0}}\xspace}

\def\Bd      {{\ensuremath{\B^0}}\xspace}

\def\BdorBdbar {\kern \thebaroffset\optbar{\kern -\thebaroffset \Bd}\xspace}
\def\Bu      {{\ensuremath{\B^+}}\xspace}
\def\Bub     {{\ensuremath{\B^-}}\xspace}
\def\Bp      {{\ensuremath{\Bu}}\xspace}
\def\Bm      {{\ensuremath{\Bub}}\xspace}
\def\Bpm     {{\ensuremath{\B^\pm}}\xspace}

\def\Bs      {{\ensuremath{\B^0_\squark}}\xspace}

\def\BsorBsbar {\kern \thebaroffset\optbar{\kern -\thebaroffset \Bs}\xspace}

\def\Y#1S{\ensuremath{\PUpsilon{(#1S)}}\xspace}

\def\LorLbar     {\kern \thebaroffset\optbar{\kern -\thebaroffset \PLambda}\xspace}

\def\to                 {\ensuremath{\rightarrow}\xspace}

\def\CP                {{\ensuremath{C\!P}}\xspace}

\def\Vud  {{\ensuremath{V_{\uquark\dquark}^{\phantom{\ast}}}}\xspace}
\def\Vcd  {{\ensuremath{V_{\cquark\dquark}^{\phantom{\ast}}}}\xspace}

\def\Vub  {{\ensuremath{V_{\uquark\bquark}^{\phantom{\ast}}}}\xspace}
\def\Vcb  {{\ensuremath{V_{\cquark\bquark}^{\phantom{\ast}}}}\xspace}

\def\AT#1     {\ensuremath{A_{\mathrm{T}}^{#1}}\xspace}           

\def\C#1      {\ensuremath{\mathcal{C}_{#1}}\xspace}                       
\def\Cp#1     {\ensuremath{\mathcal{C}_{#1}^{'}}\xspace}                    
\def\Ceff#1   {\ensuremath{\mathcal{C}_{#1}^{\mathrm{(eff)}}}\xspace}        
\def\Cpeff#1  {\ensuremath{\mathcal{C}_{#1}^{'\mathrm{(eff)}}}\xspace}       
\def\Ope#1    {\ensuremath{\mathcal{O}_{#1}}\xspace}                       
\def\Opep#1   {\ensuremath{\mathcal{O}_{#1}^{'}}\xspace}                    

\newcommand{\nospaceunit}[1]{\ensuremath{\text{#1}}}       
\newcommand{\aunit}[1]{\ensuremath{\text{\,#1}}}       

\newcommand{\tev}{\aunit{Te\kern -0.1em V}\xspace}
\newcommand{\gev}{\aunit{Ge\kern -0.1em V}\xspace}
\newcommand{\mev}{\aunit{Me\kern -0.1em V}\xspace}
\newcommand{\kev}{\aunit{ke\kern -0.1em V}\xspace}
\newcommand{\ev}{\aunit{e\kern -0.1em V}\xspace}
 
\newcommand{\mevc}{\ensuremath{\aunit{Me\kern -0.1em V\!/}c}\xspace}
\newcommand{\gevc}{\ensuremath{\aunit{Ge\kern -0.1em V\!/}c}\xspace}
\newcommand{\mevcc}{\ensuremath{\aunit{Me\kern -0.1em V\!/}c^2}\xspace}
\newcommand{\gevcc}{\ensuremath{\aunit{Ge\kern -0.1em V\!/}c^2}\xspace}

\def\mum  {\ensuremath{\,\upmu\nospaceunit{m}}\xspace}

\def\fb   {\ensuremath{\aunit{fb}}\xspace}
\def\invfb   {\ensuremath{\fb^{-1}}\xspace}

\def\deriv {\ensuremath{\mathrm{d}}}

\def\gsim{{~\raise.15em\hbox{$>$}\kern-.85em
          \lower.35em\hbox{$\sim$}~}\xspace}
\def\lsim{{~\raise.15em\hbox{$<$}\kern-.85em
          \lower.35em\hbox{$\sim$}~}\xspace}

\def\pt         {\ensuremath{p_{\mathrm{T}}}\xspace}

\def\ptot       {\ensuremath{p}\xspace}

\def\evtgen     {\mbox{\textsc{EvtGen}}\xspace}

\def\geant      {\mbox{\textsc{Geant4}}\xspace}

\def\photos     {\mbox{\textsc{Photos}}\xspace}

\def\pythia     {\mbox{\textsc{Pythia}}\xspace}

\def\tell1  {TELL1\xspace}
\def\ukl1   {UKL1\xspace}

\newcommand{\lhcborcid}[1]{\href{https://orcid.org/#1}{\hspace*{0.1em}\raisebox{-0.45ex}{\includegraphics[width=1em]{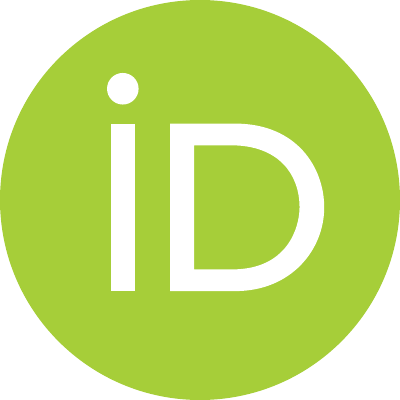}}}}

\usepackage{cite} 
\usepackage{mciteplus}

\usepackage{longtable} 
\usepackage{booktabs}
\begin{document}

\renewcommand{\thefootnote}{\fnsymbol{footnote}}
\setcounter{footnote}{1}


\begin{titlepage}
\pagenumbering{roman}

\vspace*{-1.5cm}
\centerline{\large EUROPEAN ORGANIZATION FOR NUCLEAR RESEARCH (CERN)}
\vspace*{1.5cm}
\noindent
\begin{tabular*}{\linewidth}{lc@{\extracolsep{\fill}}r@{\extracolsep{0pt}}}
\ifthenelse{\boolean{pdflatex}}
{\vspace*{-1.5cm}\mbox{\!\!\!\includegraphics[width=.14\textwidth]{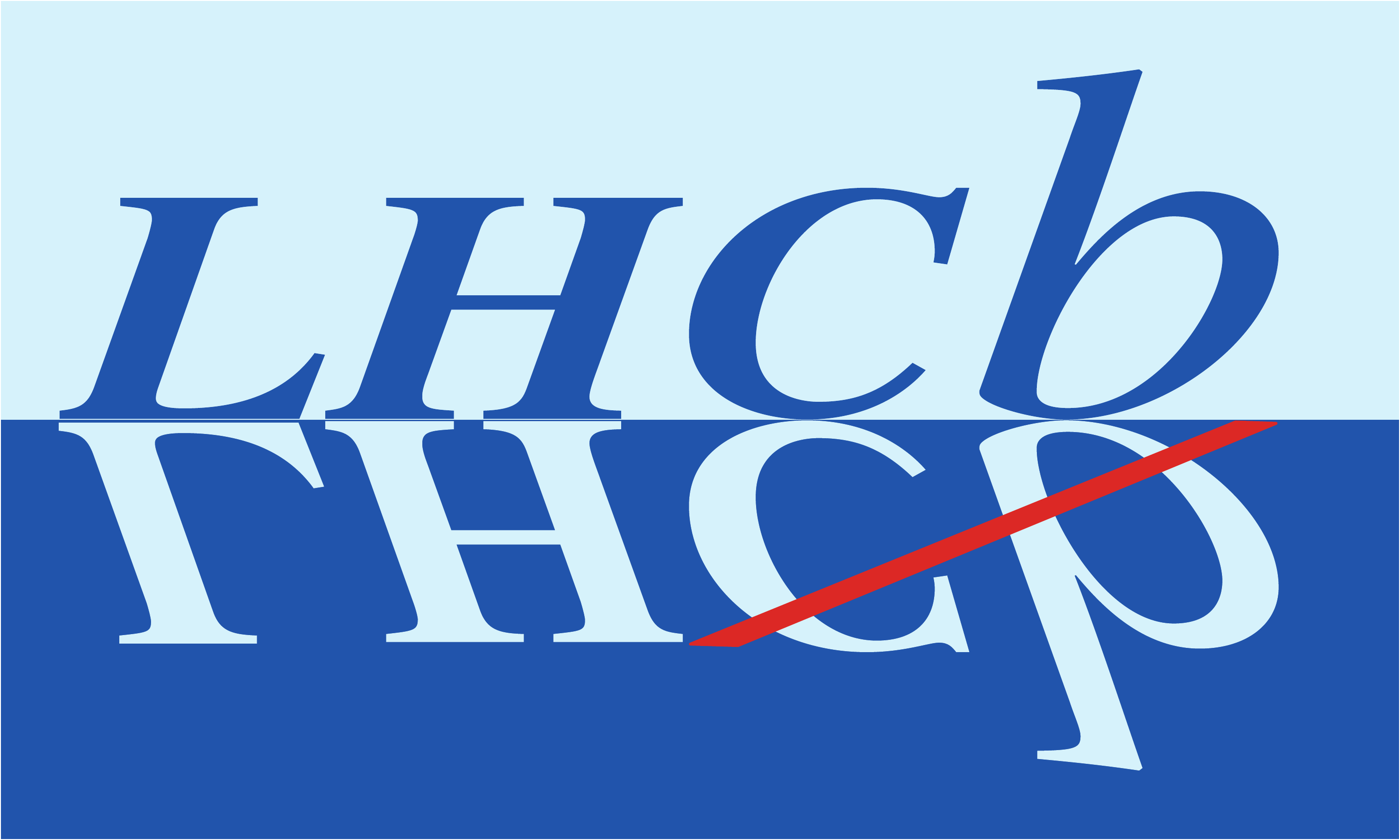}} & &}%
{\vspace*{-1.2cm}\mbox{\!\!\!\includegraphics[width=.12\textwidth]{figs/lhcb-logo.eps}} & &}%
\\  
 & & CERN-EP-2022-150 \\  
 & & LHCb-PAPER-2022-017 \\  
 & & \today \\ 
 & & \\
\end{tabular*}

\vspace*{4.0cm}

{\normalfont\bfseries\boldmath\huge
\begin{center}
  \papertitle 
\end{center}
}

\vspace*{1.5cm}

\begin{center}
\paperauthors\footnote{Authors are listed at the end of this paper.}
\end{center}

\vspace{\fill}

\begin{abstract}
  \noindent
  \noindent
  The CKM angle $\gamma$ is determined from $\CP$-violating observables measured in ${\Bpm \to D[\Kmp\pipm\pipm\pimp] h^\pm}$, $(h = K,\pi)$ decays, where the measurements are performed in bins of the decay phase-space of the $D$ meson.
   Using proton-proton collision data collected by the LHCb experiment at centre-of-mass energies of $7, 8$ and $13\tev$, corresponding to a total integrated luminosity of $9\invfb$, $\gamma$ is determined to be
  \begin{equation*}
    \gamma = \gammav,
  \end{equation*}
  where the first uncertainty is statistical, the second systematic and the third from the external inputs on the coherence factors and strong phases of the $D$-meson decays. 
\end{abstract}

\vspace*{2.0cm}

\begin{center}
  Published in JHEP 07 (2023) 138 
\end{center}

\vspace{\fill}

{\footnotesize 
\centerline{\copyright~\papercopyright. \href{\paperlicenceurl}{\paperlicence}.}}
\vspace*{2mm}

\end{titlepage}


\newpage
\setcounter{page}{2}
\mbox{~}
%
%
%
%


\renewcommand{\thefootnote}{\arabic{footnote}}
\setcounter{footnote}{0}

\cleardoublepage


\pagestyle{plain} 
\setcounter{page}{1}
\pagenumbering{arabic}


\section{Introduction}

The Standard Model (SM) description of charge-parity (\CP) violation and quark couplings in the weak interaction can be tested by measuring the parameters of the Unitarity Triangle, which is a geometrical representation of the complex plane of the Cabibbo--Kobayashi--Maskawa (CKM) quark mixing matrix\cite{Cabibbo:1963yz,Kobayashi:1973fv}.   The angle $\gamma \equiv \arg{(-{\Vud\Vub^*}/{\Vcd\Vcb^*})}$ has particular importance, as it can be determined in tree-level processes with negligible theoretical uncertainty~\cite{Brod:2013sga}. This attribute makes $\gamma$ a SM benchmark against which observables involving loop-level transitions, that are more susceptible to contributions beyond the SM, can be compared.

A powerful method by which to gain knowledge of $\gamma$ is through the measurement of \CP\ asymmetries and associated observables in $\Bpm \to D\Kpm$ decays, and related processes that involve the same quark transitions. Here $D$ indicates a \Dz\ or \Dzb\ meson reconstructed in a final state common to both, which allows for interference to occur between the CKM favoured $b \to c$ and suppressed $b \to u$ tree-level amplitudes. 
The LHCb collaboration has performed such measurements in a wide ensemble of $B$- and $D$-decay modes, giving the combined result $\gamma = \left(65.4^{+3.8}_{-4.2}\right)^\circ$~\cite{LHCb-PAPER-2021-033}. 
Final states with net strangeness, which are produced through  Cabibbo-favoured (CF) and doubly Cabibbo-suppressed (DCS) amplitudes, are of particular interest~\cite{Atwood:1996ci,Atwood:2000ck}, and the modes $D \to \Kpm\pimp$ and $D \to \Kpm\pimp\piz$ make contributions to the LHCb average~\cite{LHCb-PAPER-2021-036,LHCb-PAPER-2020-036,LHCb-PAPER-2019-021,LHCb-PAPER-2017-030,LHCb-PAPER-2015-020}. The channels $D \to  \Kpm\pimp\pimp\pipm$ belong to the same category, and are experimentally attractive due to their high branching fractions and having only charged particles in the final state. The decay rates for the four possible charge configurations are given by 
\begin{equation}
  \begin{split}
    \Gamma_{B^\pm \to D \left[ \Kmp \pipm\pipm\pimp \right] K^\pm } &\propto r_{K3\pi}^2 + (r^K_B)^2 + 2 r_{K3\pi} r^K_B R_{K3\pi} \cos(\delta^K_B + \delta_{K3\pi} \pm \gamma ) \\
    \Gamma_{B^\pm \to D \left[ \Kpm \pimp\pimp\pipm \right] K^\pm }&\propto 1 + (r_{K3\pi}^2 r^K_B)^2 + 2 r_{K3\pi} r^K_B R_{K3\pi} \cos(\delta^K_B - \delta_{K3\pi} \pm \gamma ), \\
    \end{split}
    \label{eq:nomix}
\end{equation}
where $r^K_B \approx 0.1$ is the ratio of the magnitudes of the suppressed and favoured $B$-decay amplitudes and $\delta^K_B \approx 130^\circ$ is the \CP-conserving strong-phase difference between these amplitudes~\cite{LHCb-PAPER-2021-033}.  The other parameters in Eq.~\ref{eq:nomix}, defined more precisely in the subsequent discussion, are related to the properties of the $D$-meson decay and are averaged over the inclusive multi-body phase space of the final-state particles.
The quantity $r_{K3\pi} \approx 0.06$ is the mean ratio of the DCS to the CF amplitudes and $\delta_{K3\pi} \approx 160^\circ$ is the mean strong-phase difference between these amplitudes. 
The coherence factor $R_{K3\pi} \approx 0.4 $ quantifies the dilution that the interference terms of Eq.~\ref{eq:nomix} experience from differences between the favoured and suppressed decays in the structure of the intermediate resonances~\cite{Atwood:2003mj}.  
Note that Eq.~\ref{eq:nomix} omits the small effects of charm mixing~\cite{PhysRevD.72.031501,PhysRevD.89.014021}, which will be introduced in the subsequent discussion. Furthermore, \CP\-violation in the charm system is neglected, which is an excellent approximation here~\cite{Amhis:2019ckw}.

From consideration of the form of Eq.~\ref{eq:nomix} and the size of the parameters involved, the two decays $B^\pm \to D \left[ \Kmp \pipm\pipm\pimp \right] K^\pm$ in which the two final-state kaons have opposite-sign (OS) charges are suppressed and have interference effects sensitive to $\gamma$ that appear at first order.  The two decays with kaons with like-sign (LS) charges, $B^\pm \to D \left[ \Kpm \pimp\pimp\pipm \right] K^\pm$,  are favoured and have subdominant interference effects.
The suppressed decays were first observed by LHCb and found to have a \CP\ asymmetry of $\approx -0.3$~\cite{LHCb-PAPER-2016-003}. 
The size of this asymmetry is limited by the low value of the coherence factor, which reduces the sensitivity to $\gamma$. 
Analogous expressions to Eq.~\ref{eq:nomix} can be written for the decays $B^\pm \to D \left[ \Kmp \pipm\pipm\pimp \right] \pi^\pm$ and $B^\pm \to D \left[ \Kpm \pimp\pimp\pipm \right] \pi^\pm$. 
These modes are more abundant than the $B^\pm \to D K^\pm$ modes, and have a strong-phase difference between the $B$-decay amplitudes of $\delta^\pi_B \approx 280^\circ$, but exhibit significantly lower interference effects as the value of 
$r^\pi_B$ is around $0.005$\cite{LHCb-PAPER-2021-033}.

Improved sensitivity to $\gamma$ can be achieved by studying the ${B^\pm \to D h^\pm, (h = K,\pi)}$ decay rates in separate regions, or bins, of the $D \to  \Kpm\pimp\pimp\pipm$ phase space rather than inclusively~\cite{Evans:2019wza}. 
If suitably chosen, these bins can have a higher coherence factor than that of the integrated phase space and also possess different values of the mean strong-phase difference.  
This paper reports the first use of this approach, based on a choice of four bins suggested in Ref.~\cite{Evans:2019wza} and exploiting data corresponding to an integrated luminosity of 9\invfb\ of proton-proton $(pp)$ collisions, collected by the LHCb experiment at centre-of-mass energies of 7, 8 and 13~TeV. 
Observables formed from the rates of $B$ decays are measured within each bin for both $B^\pm \to DK^\pm$ and $B^\pm \to D \pi^\pm$ decays.  These are interpreted in terms of $\gamma$ using measurements of the charm decay parameters within each bin obtained from quantum-correlated $D\Dbar$ data collected by the CLEO-c~\cite{Evans:2019wza} and BESIII~\cite{BESIII:2021eud} experiments, augmented with constraints from \Dz-\Dzb\ oscillation studies performed by the LHCb experiment~\cite{LHCb-PAPER-2015-057}.
The use of the measured $D$ parameters makes the $\gamma$ determination model independent, despite the use of $\Dz\to \Km\pip\pip\pim$ and $\Dz\to\Kp\pim\pim\pip$ amplitude models  ~\cite{LHCb-PAPER-2017-040} to define the phase-space bins.\footnote{The inclusion of charge-conjugate processes is implied throughout, except in discussions of asymmetries.}
\FloatBarrier
\section{Bin definitions and observables} 

The coherence factor $\Cf$ and the average strong-phase difference $\sp$ are defined by 
\begin{equation}
  \Cf e^{i \sp } \equiv \left( A_{\Dz} A_\Dzb \right)^{-1} \int \deriv{\psi} \mathcal{A}_{\Dzb  } ( \psi ) \mathcal{A}^{\ast}_{ \Dz } ( \psi ),  
  \label{eq:coherenceFactor}
\end{equation} 
where $\mathcal{A}_{\Dz}(\psi)$ and $\mathcal{A}_{\Dzb}(\psi)$ are the DCS and CF decay amplitudes to the $\Kp\pim\pim\pip$ final state as a function of the position in the phase space $\psi$. The phase-space density is by convention included in the definition of the infinitesimal volume element $\deriv{\psi}$.  
The normalisation integrals $A_{\Dz(\Dzb)}$ are given by 
\begin{equation}
  A_{\Dz(\Dzb)} = \sqrt{ \int \text{d} \psi \left| \mathcal{A}_{\Dz(\Dzb)}(\psi) \right|^2 }.
\end{equation}
When many intermediate states with differing amplitudes contribute to the $\Dz$ and $\Dzb$ decays, 
a wide range of phase differences in the integrand of Eq.~\ref{eq:coherenceFactor}
leads to a low coherence factor. 
Conversely, decays have higher coherence when a single intermediate state dominates both transitions.
This suggests a strategy to partition the phase space:
dividing it into regions where the range of phase differences is smaller than over the entire phase space will result in higher coherence and thus improve the sensitivity to $\CP$ violation. 
The scheme  used in this paper was proposed in Ref.~\cite{Evans:2019wza} and splits the decays into four bins using the {\it normalised strong-phase difference}, defined as 
\begin{equation}
    \tilde{\delta}_{K3\pi}(\psi) \equiv \mathrm{arg}\left( \mathcal{A}_{\Dzb}(\psi) \mathcal{A}^{\ast}_{\Dz}(\psi) \right) 
                                - \mathrm{arg}\left( \int  \text{d} \psi^\prime \mathcal{A}_{\Dz}(\psi^\prime) \mathcal{A}^{\ast}_{\Dzb}(\psi^\prime) \right),
\end{equation}
where the amplitudes $\mathcal{A}_{\Dz(\Dzb)}(\psi)$ are evaluated using the models of Ref.~\cite{LHCb-PAPER-2017-040}. 
The second term ensures that the average is zero, as the models are insensitive to the absolute phase difference between the two amplitudes. 
A fifth veto bin, orthogonal to the other four, is defined to capture the decays $\Dz \to \KS[
\pip\pim]\Kpm \pipm$, which are difficult to separate topologically from the $\Dz\to \Km\pip\pip\pim$ final state when the $\Dz$ meson is produced at rest, as is the case at the CLEO-c and BESIII experiments. 
This bin is defined by either opposite-sign dipion pair being within 10\mevcc of the known \KS mass \cite{PDG2022}, and removes around $5\%$ of signal candidates. 
The number of $B\to D h$ decays in each bin is proportional to the rate given in Eq.~\ref{eq:nomix}, but with local coherence factors, denoted by $R^i_{K3\pi}$, that are higher than the phase-space integrated value.
Measurements of the hadronic parameters have been performed in the four bins motivated by the $\psi(3770)$ data from the CLEO-c experiment~\cite{Evans:2019wza}, and augmented with results from a larger data set collected by the BESIII collaboration~\cite{BESIII:2021eud}. 
The combined results from these analyses, which also benefit from a study of charm mixing performed by the LHCb collaboration\cite{LHCb-PAPER-2015-057}, are reported in Table~\ref{bin_definitions}, together with definitions of the four bins.

\begin{table}
  \caption{\label{bin_definitions}Definition of the four bins in terms of $\tilde{\delta}_{K3\pi}$. The measured values of $R_{K3\pi}^i$ and $\delta_{K3\pi}^i$, where $i$ refers to the phase-space bin, are taken from Ref.~\cite{BESIII:2021eud}, and combine measurements from BESIII, CLEO-c and an LHCb analysis of charm mixing. 
  }
  \centering
    \def\arraystretch{1.2}
\begin{tabular}{lrrr}
\toprule
  Bin & Limits $(\tilde{\delta}_{K3\pi})$ 
    & \multicolumn{1}{c}{$R^{i}_{K3\pi}$}
    & \multicolumn{1}{c}{${\delta^{i}_{K3\pi}}$}\\
\midrule
  1 & $          -180^\circ < \tilde{\delta}_{K3\pi} \leq -\phantom{1}39^\circ $ & $0.66 \aerr{0.18}{0.21}$ & $\left(117\aerr{14}{19}\right)^\circ $ \\
  2 & $-\phantom{1}39^\circ < \tilde{\delta}_{K3\pi} \leq \phantom{-10}0^\circ $ & $0.85 \aerr{0.14}{0.21}$ & $\left(145\aerr{23}{14}\right)^\circ $ \\
  3 & $\phantom{-18}0^\circ < \tilde{\delta}_{K3\pi} \leq \phantom{-0}43^\circ $ & $0.78 \aerr{0.12}{0.12}$ & $\left(160\aerr{19}{20}\right)^\circ $ \\
  4 & $\phantom{-1}43^\circ < \tilde{\delta}_{K3\pi} \leq \phantom{-}180^\circ $ & $0.25 \aerr{0.16}{0.25}$ & $\left(288\aerr{15}{29}\right)^\circ $ \\
\bottomrule
\end{tabular}
\end{table}

The observables used to determine $\gamma$ and related hadronic parameters are the ratios of rates of OS-to-LS $\Bpm\to D h^\pm$ decays in each phase-space bin. These observables are given in the $i$th bin by
\begin{equation}
  \begin{split}
    \mathcal{R}^{i}_{h^\pm} = \biggl(& (\ri)^2 + (r_B^h)^2 + 2 \ri r^h_B \cfi \cos(\delta^h_B+\spi\pm\gamma)                                                       \\ 
                    &- \ri \cfi \left( y \cos\spi - x \sin\spi \right)     + \frac{1}{2} \left( x^2 + y^2\right)    \\
                    &- r_B^h \left( y \cos(\delta_B^h\pm\gamma) + x \sin(\delta_B^h\pm\gamma) \right) \biggr)  \\ 
                    & \bigg/ \biggl( 1 + (r_B^h)^2 (\ri)^2 + 2 \ri \cfi r_B^h \cos(\delta_B^h -\spi \pm \gamma) \biggr),
    \label{eq:master}
  \end{split}
\end{equation}
where the average ratio of $D$-decay amplitudes is denoted by $r^i_{K3\pi}$. 
The effects of charm mixing are now included, in contrast with Eq.~\ref{eq:nomix}. 
This is governed by the parameters $x$ and $y$, both of which are smaller than $1\%$~\cite{Lenz:2020awd,Amhis:2019ckw}.  

A complementary set of observables integrated over the phase space, including the $\KS$-veto bin, is also reported.
These observables are the decay asymmetry $\mathcal{A}_{h}$, defined as the ratio of the difference in $\Bm$ and $\Bp$ yields to their sum,
and the charge-averaged OS-to-LS ratio, denoted by $\mathcal{R}_{h}$. 
These inclusive observables allow for comparison with the results of previous studies~\cite{LHCb-PAPER-2016-003}. 
The decay asymmetry is also reported in each phase-space bin, as it is expected to be approximately proportional to $\sin(\delta_B^{h} + \delta^i_{K3\pi})$, and therefore has a more intuitive evolution with the strong-phase difference than the yield-ratio observables.  

Additional observables are provided by flavour-tagged $\Dz \to K^\pm\pi^\mp\pi^\mp\pi^\pm$ decays, which are produced in the decays ${X_b \to D^{*+}[\Dz \pip] \mu^{-} \neumb X}$, where $X$ can be several additional particles that are not reconstructed and $X_b$ is any hadron containing a $\bquark$ quark that decays to this final state. 
This data set is referred to as doubly tagged (DT), as the charges of both the pion and muon give the flavour of the $D$ hadron at its production. 
The ratios of yields of decays where the pion from the $D^{*+}$ decay is of the same charge as the kaon to those where they are of opposite charge are given by 
\begin{equation}
  \mathcal{R}_{\text{DT}}^i = (\ri)^2 - \ri \cfi \left( y \cos\spi - x \sin\spi \right) + \frac{1}{2} \left( x^2 + y^2\right),
\end{equation}
and are included as additional constraints in the fit to $\gamma$ and the associated hadronic parameters. 
These observables constrain the \ri~parameters, which consequently improves the precision of the $B\to D\pi$ strong-phase difference by around $30\%$.

\FloatBarrier 

\section{The LHCb detector}

The \lhcb detector~\cite{LHCb-DP-2008-001,LHCb-DP-2014-002} is a single-arm forward
spectrometer covering the \mbox{pseudorapidity} range $2<\eta <5$,
designed for the study of particles containing \bquark or \cquark
quarks. The detector includes a high-precision tracking system
consisting of a silicon-strip vertex detector surrounding the $pp$
interaction region~\cite{LHCb-DP-2014-001}, a large-area silicon-strip detector located
upstream of a dipole magnet with a bending power of about
$4{\mathrm{\,Tm}}$, and three stations of silicon-strip detectors and straw
drift tubes~\cite{LHCb-DP-2017-001}
placed downstream of the magnet.
The tracking system provides a measurement of the momentum, \ptot, of charged particles with
a relative uncertainty that varies from 0.5\% at low momentum to 1.0\% at 200\gevc.
The minimum distance of a track to a primary $pp$ collision vertex (PV), the impact parameter, 
is measured with a resolution of $(15+29/\pt)\mum$,
where \pt is the component of the momentum transverse to the beam, in\,\gevc.
Different types of charged hadrons are distinguished using information
from two ring-imaging Cherenkov detectors~\cite{LHCb-DP-2012-003}. 
Photons, electrons and hadrons are identified by a calorimeter system consisting of
scintillating-pad and preshower detectors, an electromagnetic
and a hadronic calorimeter~\cite{LHCb-DP-2020-001}. Muons are identified by a
system composed of alternating layers of iron and multiwire
proportional chambers~\cite{LHCb-DP-2012-002}.
The online event selection is performed by a trigger~\cite{LHCb-DP-2012-004, LHCb-DP-2019-001}, 
which consists of a hardware stage, based on information from the calorimeter and muon
systems, followed by a software stage, which applies a full event
reconstruction.

Simulation is required to model the effects of the detector acceptance and the
imposed selection requirements.
In the simulation, $pp$ collisions are generated using
\pythia~8~\cite{Sjostrand:2006za, Sjostrand:2007gs} 
with a specific \lhcb configuration~\cite{LHCb-PROC-2010-056}.
Decays of unstable particles
are described by \evtgen~\cite{Lange:2001uf}, in which final-state
radiation is generated using \photos~\cite{davidson2015photos}.
The interaction of the generated particles with the detector, and its response,
are implemented using the \geant
toolkit~\cite{Allison:2006ve, *Agostinelli:2002hh} as described in
Ref.~\cite{LHCb-PROC-2011-006}. 
The underlying $pp$ interaction is reused multiple times, with an independently generated signal decay for each~\cite{LHCb-DP-2018-004}.

\section{Data set and candidate selection}

The analysis uses data collected by the \lhcb experiment in $pp$ collisions, corresponding
to integrated luminosities of $1\invfb$, $2\invfb$ and $6\invfb$ collected at $\sqrt{s} = 7, 8$ and $13\tev$, respectively. 
Signal candidates are reconstructed by pairing a four-body $D$-meson candidate, with reasonable separation from any PV, with an additional particle referred to as \textit{companion} hadron.
The $B$ candidate is required to have a well-separated decay vertex and approximately point to a PV.
The four possible combinations of relative charges (OS, LS) and companion-hadron flavour ($K, \pi$) are all selected identically such that the efficiency corrections to the observables are small. 
After candidates are reconstructed, further selection is required to suppress both combinatorial background and contributions from specific processes, such as when one or more tracks are assigned an incorrect mass hypothesis.

The combinatorial background, which is composed of both real $D$ mesons paired with an additional random track and also $D$ candidates that contain at least one track from the rest of the event, are suppressed using a deep neural network (DNN)~\cite{Hocker:2007ht}. 
The DNN is trained using simulated $B^\pm\to D \pi^\pm$ decays as a proxy for the signal and as background $\Bpm\to D \pi^\pm$ candidates where the mass of the $B$-meson candidate is above $5.9\gevcc$. 
The DNN uses kinematic and topological variables of the $B$ and $D$ decays, such as the displacement between the primary and decay vertices or the transverse momentum of the companion hadron.
To minimise the variation in the acceptance across the phase space, the momenta of $D$-decay products are not used.
The requirement on the output of the classifier is optimised to minimise the uncertainty on the angle $\gamma$, which is estimated by performing fits to ensembles of simulated data sets. 

After applying the DNN selection, further requirements are made to suppress specific sources of background. 
The contribution from misidentified $\Bpm\to D\pipm$ decays in the $\Bpm \to D \Kpm$ sample is suppressed by requiring that it is highly probable, according to the corresponding particle identification (PID) variable, that the companion hadron is a kaon as opposed to a pion. 
Candidates that fail this requirement are selected as $\Bpm\to D\pipm$ decays, such that the two samples are mutually exclusive. 
Candidates that do not contain a correctly reconstructed charm hadron are suppressed by requiring that the mass of the $D$ meson is within $18\mevcc$ of its known value \cite{PDG2022}. 
The $D$ decay vertex is required to be downstream of the $B$ decay vertex by at least twice the uncertainty on the difference between the longitudinal vertex positions,
which in addition to suppressing the combinatorial background largely removes charmless $B$ decays such as $\Bp \to \Km\Kp\pip\pim\pip$. 

The minimum opening angle between pairs of decay products from the $D$ candidate, evaluated in the lab frame, must be greater than $0.03^\circ$
to remove candidates where one of the tracks has been erroneously duplicated in the event reconstruction. 
Additional sources of background come from $B \to D h h^\prime h^{\prime\prime}$ decays, where $h^{(\prime)(\prime\prime)}$ denotes either a pion or a kaon and the $D$ meson decays to a two-body final state. 
Such processes can be reconstructed in the $B\to D[\Km\pip\pip\pim] h$ final state if 
some of the $D$ decay products have actually come from the $B$ decay, and vice versa. 
This source of background is suppressed by the previously described requirements that select well reconstructed $D$-meson candidates.  
The residual contamination from such decays is removed by requiring the minimum combined mass of the companion hadron and one of the $D$-decay products is greater than 15\mevcc from the known mass of the $D$ meson; only combinations that are CKM favoured with respect to the OS decays, such as $\Bp \to \PD\left[\Km\Kp\right]\pip\pim\pip$, are considered

Lastly, LS decays can be misidentified as an OS decay via a misidentification of both the kaon and one of the pions from the $D$ decay. 
This source of background is referred to as \textit{crossfeed}, and is suppressed by the requirements on the PID of the $D$-decay products. 
In addition, $D$ candidates are vetoed if they fall within 15\mevcc of the known $\Dz$ mass after interchanging the kaon and pion mass hypotheses. 
The rate of the residual contamination from this background is estimated to be $(1.02\pm0.14) \times 10^{-4}$, which corresponds to around 100 such decays in the OS $\Bpm \to D \pipm$ sample or around $3\%$ of the anticipated signal yield. 

\section{Mass fit} 
\label{sec:massFit}
Fits are performed on the mass distribution of the selected $B$ candidates to determine the signal yields, the contributions of the various sources of background, and their corresponding uncertainties.
The mass of the candidate $B$ meson, denoted by $m_{Dh}$, is calculated with the $D$ candidate constrained to its known mass~\cite{PDG2022} and the $B$ constrained to originate from a primary vertex~\cite{Hulsbergen:2005pu}. 
A likelihood fit, binned in mass with intervals of $1\mevcc$, is first performed integrating over the phase space, in order to ascertain the quality of the signal and background description and to measure the yields for the inclusive analysis.
The mass fit is performed in the range $5.1$ to $5.9\gevcc$, where the low mass limit is used to suppress the more complex background contributions that are present at lower masses. 
The phase-space integrated fit is performed on eight subsamples simultaneously, where the samples are divided by: the flavour of the companion hadron; the charge of the $B$ candidate; and  the charge of the kaon from the $D$-meson decay relative to that of the $B$ candidate.
The fit is then repeated, splitting the candidates into the bins of the phase space, leading to a total of 32 samples. The bin of each candidate is determined constraining the $D$ candidate to its known mass \cite{PDG2022}, ensuring that all reconstructed decays are within the same well-defined phase space. 

The yields of the LS $\Bpm\to D\pipm$ candidates in each phase-space bin are allowed to freely vary. 
The ratio of yields of LS $\Bpm \to \PD \Kpm$ to $ \Bpm \to \PD \pipm$ candidates in each bin is also allowed to vary freely.
The yields of the OS decays are then related to the corresponding LS yields via the observables $\mathcal{R}^i_{h^\pm}$.
A candidate can be assigned to an incorrect phase-space bin due to the finite resolution of the detector.
This effect is found to be small, with more than 95\% of candidates reconstructed in the correct bin. 
The migration between bins is corrected using simulation, by evaluating the probabilities that a candidate is reconstructed in each of the bins, given the known true bin. 
These probabilities are then included in the mass fit as a migration matrix. 
The difference in reconstruction and selection efficiencies between the OS and LS decays depend weakly on the distributions of the candidates within each bin, and therefore also on the physics parameters, in particular due to charm mixing. 
Corrections for such effects are applied in the interpretation fit described in Sect.~\ref{sec:interpretationFit}. 

The signal mass distribution is described by a Gaussian distribution modified to accommodate both an asymmetric shape and broader tails, 
\begin{equation}
  \mathcal{P}_\text{sig.}(m) \propto \exp\left( \frac{ -( m -\mu)^2  \left( 1 + \alpha (m-\mu)^2 / (2 \sigma_{w}^2) \right) }{ 2 \sigma^2 + \alpha(m-\mu)^2 } \right),
  \label{eq:cruijff}
\end{equation}
where $\alpha$ and $\sigma$ have different values on either side of the most probable value of the mass, $\mu$. 
The second term of the numerator ensures that the distribution tends to a Gaussian at $|x - \mu| \gg \sigma$, with the width of this Gaussian given by $\sigma_{w} \sim 200\mevcc$. 
The $\alpha$ and right-handed $\sigma$ parameters of the signal distribution of the $B^\pm \to D\pi^\pm$ contribution are allowed to vary in the fit to data, while all others are fixed with respect to these two parameters and the simulation. 
The value of $\mu$ is also allowed to vary in the fit and have a different value for the $\Bp$ and $\Bm$ distributions, due to small differences in the detector response to the different charges. 

A prominent source of background that peaks close to the signal arises from decays where the companion hadron has been misidentified, with $\Bpm\to D\pipm$ decays peaking around $40\mevcc$ above the $B$ mass in the
${\Bpm \to D \Kpm}$ 
sample, and $\Bpm \to D \Kpm$ decays around $25\mevcc$ lower in the $\Bpm\to D\pipm$ sample. 
The shape of this background is modelled with a double-sided Crystal-Ball~\cite{Skwarnicki:1986xj} distribution, while the yield is fixed with respect to the corresponding correctly reconstructed decay after accounting for the relative efficiencies of the PID requirements from data~\cite{LHCb-PUB-2016-021}. 

The partially reconstructed background decays where the intermediate state is of negligible natural width, such as $\Bp \to D^{*}\left[\PD\piz\right] h^+$, are modelled by convolving a low-order polynomial function with the same mass resolution function used to describe the fully reconstructed decay. The parameters of polynomial function depend on the orbital structure and available phase space of the decay~\cite{LHCb-PAPER-2020-036}, and are determined from simulation. 
The yields of such background sources in the LS sample are constrained relative to the fully reconstructed mode using relative branching ratios and efficiencies. 
The yields in the OS samples, both integrated and in the phase-space bins, are fixed with respect to the corresponding LS yields. 

For partially reconstructed decays where the intermediate state is broad, such as 
${\Bp \to \PD {\rho^+}\left[\pi^+ \piz\right]}$, the distribution is described by the convolution of two polynomial functions with a pair of Gaussian distributions.
The shapes of these processes are determined from simulation, while the yields are treated identically to the partially reconstructed decays with narrow intermediate states. 
A closely related category of decays are those where an additional charged particle is missed. These decays are particularly prominent in the OS samples, specifically from $\Bs\to \Dzb \Kp \pim$ and $\Bz\to \Dzb\pip\pim$ decays in the $\Bpm \to D \Kpm$ and $\Bpm \to D \pipm $ samples, respectively. 
The shapes of these backgrounds are taken from the dedicated studies of each of these three-body decays performed by the LHCb collaboration~\cite{LHCb-PAPER-2014-036, LHCb-PAPER-2014-070}. The yields of these background sources vary freely in the fit, and are distributed in the phase space identically to LS decays.

The last category of partially reconstructed decays are those largely removed by the $m_{Dh} > 5.1\gevcc$ requirement, and includes contributions from heavier $D^*$ resonances such as the $D^*(2300)$ meson, or decays where multiple particles are not reconstructed, such as $\Bpm \to D a_1^\pm$. 
The distribution of such background sources are fixed using simulation, while yields are fixed using branching ratios with respect to the fully reconstructed LS $\Bpm \to D \pipm$ mode and relative efficiencies. 

Background candidates composed of random combinations of particles can be divided into two categories. 
In the first, a correctly reconstructed $D$ meson is paired with a random hadron. 
Such candidates are distributed exponentially in $m_{Dh}$, while in the $D$ phase space the distribution is that of the Cabibbo-favoured signal. 
In the second category, at least one of the tracks from the $D$ candidate is produced by another process, most often from the primary $pp$ interaction.
The distribution in $m_{Dh}$ and in the $D$ phase space can be assessed by studying the candidates that fail the $D$ mass requirement. 
The yields in each bin in the phase space are allowed to vary under constraint in the signal fit, and are approximately distributed according to the four-body phase-space density and have negligible $\CP$ violation. 
The shape in $m_{Dh}$ is well described by the exponentiation of a third-order Chebychev polynomial function, and is found to have indistinguishable distributions between the different phase-space bins. 

The crossfeed background from the LS mode into the OS sample is modelled using the same shape as the signal, Eq.~\ref{eq:cruijff}, but with all parameters fixed to those obtained from a fit to the simulated LS decays that have been reconstructed in the OS sample. 
The distribution in phase space is approximately the same as the CF decays, with corrections derived from the aforementioned simulated sample of LS decays used to account for the additional migration of candidates caused by the incorrect mass hypotheses for two of the $D$-decay products. 

\section{Systematic uncertainties}

Systematic uncertainties arise from a number of effects that could affect the measurements of the $\mathcal{R}^i_{h^\pm}$ observables.
The total systematic uncertainty is obtained by taking the sum of all contributions in quadrature, and is an order of magnitude smaller than the statistical uncertainty. The dominant contributions come from corrections to the signal yields, such as from detection asymmetries.  

The probability of placing a candidate where the companion particle is truly a kaon in the $\Bm \to D \Km$ sample, rather than the $\Bm \to D \pim$ sample, is evaluated using data \cite{LHCb-PUB-2016-021}, and is found to be $(65.5\pm 0.7)\%$, where the uncertainty is estimated by varying the binning scheme used to calculate the efficiency.
The corresponding probability for the companion pion is found to be above $99\%$, but is allowed to vary in the fit. 
The yield observables are affected by correlations between the efficiency of the particle identification requirements on the $D$-decay products and their kinematics.
The effect is very small, and thus only applied as a source of systematic uncertainty, which is evaluated by counting the number of $\Bpm\to D\pipm$ LS candidates in each bin before and after applying these requirements. The uncertainty is taken as the maximal relative change between the bins, which is found to be around $0.5\%$. This source of systematic uncertainty also affects the DT sample. 
The uncertainty on the bin migration is taken as half of the value of the correction from simulation. 
This source of systematic uncertainty also affects the DT sample. 
The detection efficiencies of particles and antiparticles differ due to the difference in interaction cross-sections with the detector material.
The asymmetry for pions is small, with a value of $\left(-0.17\pm0.10\right)\%$, while the asymmetry for kaons is larger, and found to be $\left(-0.79\pm0.25\right)\%$. 
These values are taken from dedicated studies \cite{LHCb-PAPER-2016-054, LHCb-PUB-2018-004}, correcting for the kinematics of the decays of interest. 
The production asymmetry is a free parameter of the fit, determined by assuming that \CP violation is negligible in the  LS $\Bpm \to D \pipm$ decays, and thus no additional systematic uncertainty is required. 
   
There are several sources of systematic uncertainty related to the partially reconstructed background. 
First, there are the uncertainties in the distributions of each type of background, which are assessed by varying within their uncertainties the shapes obtained from simulation. 
Second, the yields of some background processes are fixed relative to the fully reconstructed LS $B^\pm \to D \pipm$ decay, and thus the knowledge of relative branching fractions and the finite size of simulated samples used to assess the relative efficiencies of these modes are both sources of systematic uncertainty. 
Last, the $\CP$ violation in some of the suppressed partially reconstructed decays, such as $\Bpm\to D^{*} \Kpm$, is expected to be significant. 
The \CP violation in the partially reconstructed background is accounted for by fixing the yields of the suppressed, partially reconstructed decays to their expected values based on the current knowledge of $\gamma$ and the hadronic parameters, with values taken from Ref.~\cite{LHCb-PAPER-2021-033}. The corresponding uncertainty is propagated as a source of systematic uncertainty. 

The final sources of systematic uncertainty are due to the modelling of the other sources of background. This includes the shapes of the different combinatorial components, in addition to the yields and shapes of the favoured-to-suppressed crossfeeds. A small number of $\Bpm\to\Kp\Km\pip\pim\pipm$ decays are also expected in the OS $\Bpm\to D\Kpm$ sample. The contamination from this background is estimated using the candidates that fail the requirements on the $\Dz$ candidate, and is found to be around two. The number of candidates, distribution in phase space and possible \CP violation in this component are all varied as a source of systematic uncertainty. 

\FloatBarrier
\section{Results}

\begin{figure}
  \includegraphics[width=0.939\textwidth]{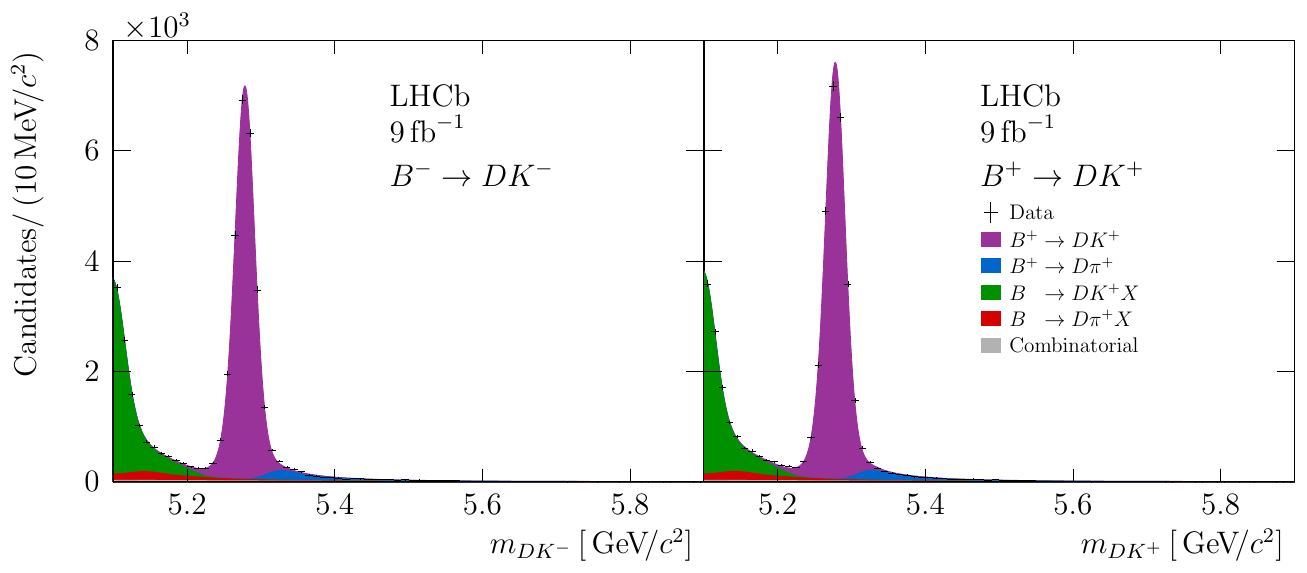}
  
  \includegraphics[width=0.939\textwidth]{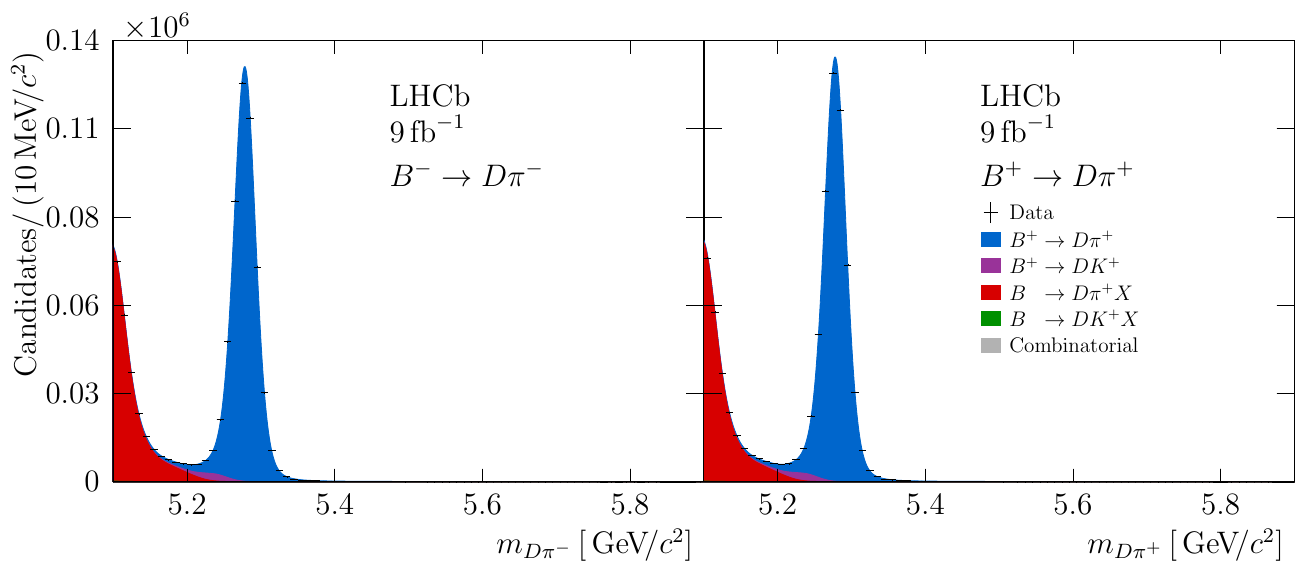}
 
  \caption{\label{massFit_integrated_LS}Invariant-mass distributions of LS $\Bpm \to D \Kpm$ (top) and $\Bpm \to D \pipm$ (bottom) candidates, divided by charge of the $B$ hadron. The results of the fit are overlaid.  } 
\end{figure}

\begin{figure}
  
  \includegraphics[width=0.939\textwidth]{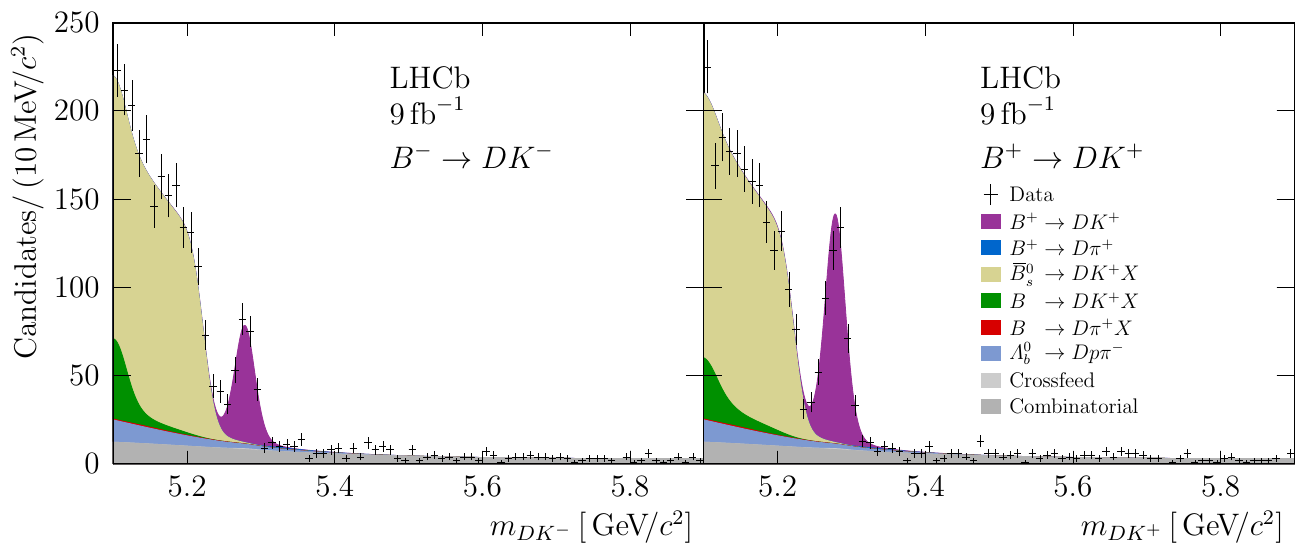}
  
  \includegraphics[width=0.939\textwidth]{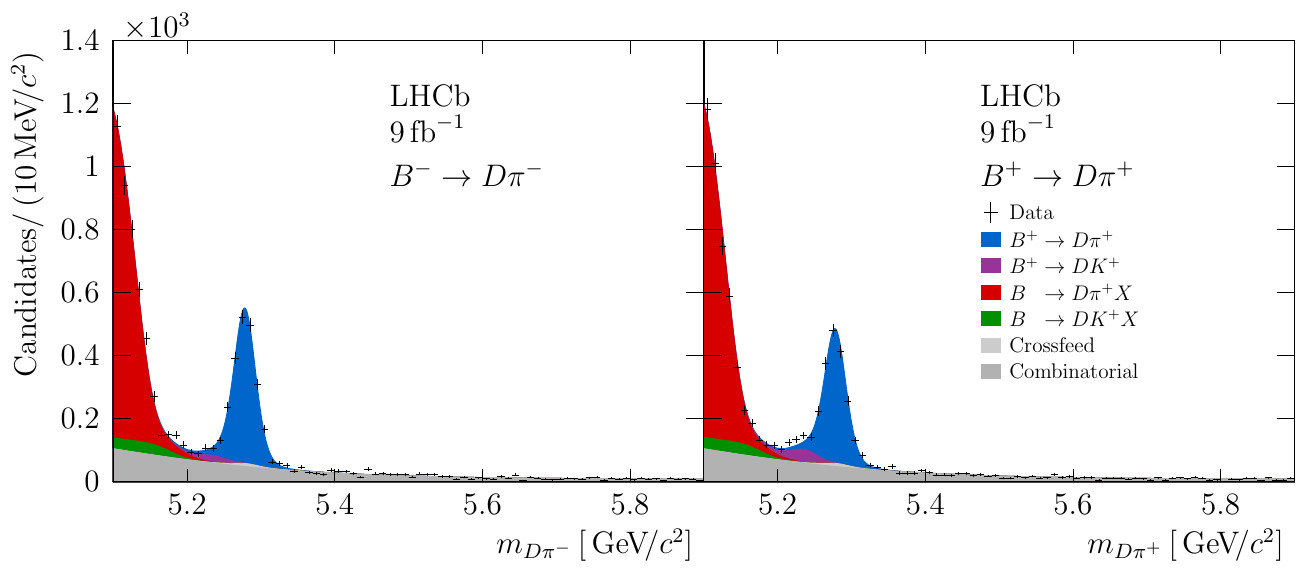}

  \caption{\label{massFit_integrated_OS} Invariant-mass distributions of OS $\Bpm \to D \Kpm$ (top) and $\Bpm \to D \pipm$ (bottom) candidates, divided by charge of the $B$ hadron. The results of the fit are overlaid.   } 
\end{figure}

The phase-space integrated fits to the eight different signal categories are shown in Figs.~\ref{massFit_integrated_LS} and~\ref{massFit_integrated_OS} for the LS and OS subsamples, respectively.
The model is found to describe the data very well, with a $\chi^2$ per degree-of-freedom of $1228 / 1249 $, corresponding to a $p$-value of $66\%$.
The phase-space integrated $\CP$ asymmetry $\mathcal{A}_h$ and the charge-averaged decay rate of the OS with respect to the LS mode, $\mathcal{R}_h$, are found to be 
\begin{equation*}
  \begin{split}
    \mathcal{A}_{K}   &=           -0.321 \pm 0.039 \pm 0.005, \\
    \mathcal{A}_{\pi} &= \phantom{-}0.070 \pm 0.019 \pm 0.006, \\
    \mathcal{R}_{K}   &= \left(  13.33 \pm 0.55 \pm 0.08\right) \times 10^{-3}, \\
    \mathcal{R}_{\pi} &= \left( \phantom{1}3.45 \pm 0.07 \pm 0.01\right) \times 10^{-3}, \\
  \end{split}
\end{equation*}
where the first and second uncertainties are statistical and systematic, respectively. 
The values are consistent with those obtained by the phase-space integrated analysis~\cite{LHCb-PAPER-2016-003} performed on the 7 and 8 $\tev$ data sets recorded in 2011 and 2012, with uncertainties that are around a factor of $2.5$ smaller. 

\begin{figure}
  \includegraphics[width=0.939\textwidth]{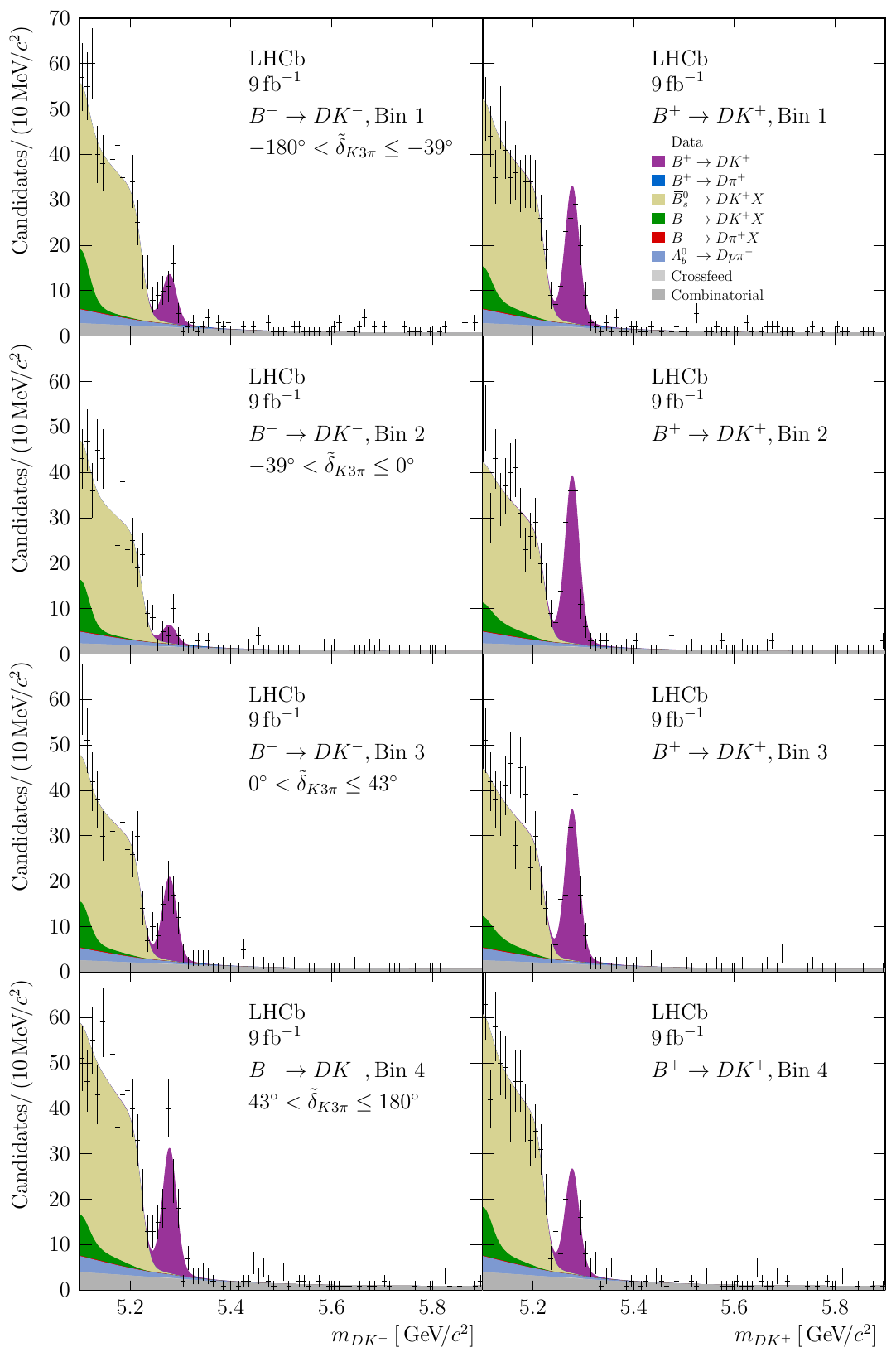}
  \caption{\label{massFit_DK}Invariant-mass distributions of OS $\Bpm \to D \Kpm$ candidates, divided by the charge of the $B$-hadron and phase-space bin. The results of the fit are overlaid.  }
\end{figure}

\begin{figure}
  \includegraphics[width=0.939\textwidth]{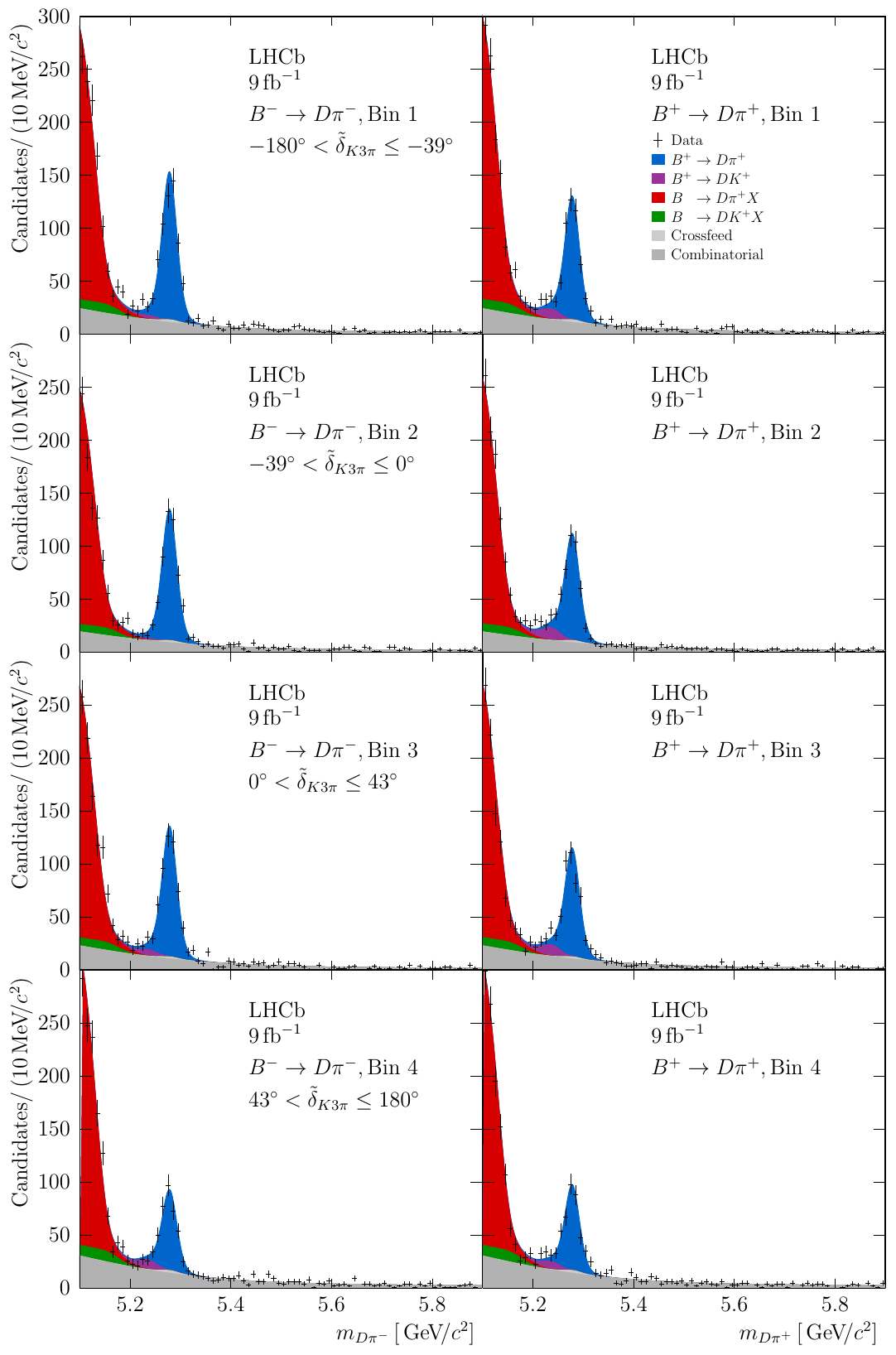}
  \caption{\label{massFit_Dpi}Invariant-mass distributions of OS $\Bpm \to D \pipm$ candidates, divided by the charge of the $B$-hadron and phase-space bin. The results of the fit are overlaid. }
\end{figure}

The data set is then divided into the different bins of phase space and the mass fit repeated. 
The fit to the 16 different suppressed signal categories is shown in Figs.~\ref{massFit_DK} and ~\ref{massFit_Dpi} for  the $\Bpm\to D\Kpm$ and $\Bpm\to D\pipm$ samples, respectively. 
Large $\CP$ violation is observed in three of the four $D$ phase-space bins in the $\Bpm\to D\Kpm$ sample.
The $\CP$ asymmetries for the kaon observables in the four bins are 
\begin{equation*}
  \begin{split}
    \mathcal{A}^{1}_{K} &= -0.469 \pm 0.088 \pm 0.009, \\
    \mathcal{A}^{2}_{K} &= -0.852 \pm 0.077 \pm 0.012, \\
    \mathcal{A}^{3}_{K} &= -0.284 \pm 0.080 \pm 0.009, \\
    \mathcal{A}^{4}_{K} &= +0.107 \pm 0.083 \pm 0.009, \\
  \end{split}
\end{equation*}
where the first and second uncertainties are statistical and systematic, respectively. 
The first three asymmetries are individually significant, with the $\CP$ violation in the second bin alone at over 10 standard deviations, and a central value larger in magnitude than any other $\CP$ asymmetry yet observed. 
The $\CP$ asymmetry in the fourth bin is not found to be statistically significant, and is expected to be smaller than the asymmetry in the other bins due to the lower amplitude ratio and coherence factor. 
The corresponding asymmetries for the pion mode are 
\begin{equation*}
  \begin{split}
    \mathcal{A}^{1}_{\pi} &= +0.087 \pm 0.037 \pm 0.006, \\
    \mathcal{A}^{2}_{\pi} &= +0.101 \pm 0.040 \pm 0.006, \\
    \mathcal{A}^{3}_{\pi} &= +0.090 \pm 0.040 \pm 0.006, \\
    \mathcal{A}^{4}_{\pi} &= -0.041 \pm 0.054 \pm 0.007. \\
  \end{split}
\end{equation*}
The significances in the first three bins are between two and three standard deviations, while the fourth bin is found to be compatible with \CP conservation. The $p$-value of the $\Bpm \to D \pipm$ observables under the $\CP$-conserving hypothesis when combining the bins is less than $2\times 10^{-3}$. The $\CP$ asymmetries are of the opposite sign to those in the kaon mode, which is consistent with the strong-phase difference for the $\Bpm \to D\pipm$ amplitudes being separated by roughly $180^\circ$ from that of the $\Bpm \to D \Kpm$ amplitudes. 
The yield ratio observables, and corresponding uncertainties, are given in Table~\ref{obs_table}. 
The observables and corresponding uncertainties for the doubly-tagged data set are also given in Table~\ref{obs_table}, while details of the selection and mass fits used to determine these values are given in App.~\ref{app:DT}. 

\begin{table}
  \caption{\label{obs_table}Observables $\mathcal{R}_{h^\pm}$ and $\mathcal{R}_\text{DT}$ in the four phase-space bins. Values are presented without correcting for differences in the efficiencies between OS and LS decays. Corrections of $\mathcal{O}(1)\%$ are applied in the interpretation fit, and are given in App.~\ref{app:covAndEfficiency}. }
  \centering
  \def\arraystretch{1.2}
  \scalebox{0.85}{
  \begin{tabular}{l | ccccc}
    \toprule
    Bin & $\mathcal{R}_{\Km} \left[\times10^{-3}\right]$  & $\mathcal{R}_{\Kp}\left[\times10^{-3}\right] $ & $\mathcal{R}_{\pim}\left[\times10^{-3}\right]$ & $\mathcal{R}_{\pip}\left[\times10^{-3}\right]$ & 
    $\mathcal{R}_{\text{DT}}\left[\times10^{-3}\right]$ \\ 
    \midrule 
    1   & $\phantom{1}5.89 \aerr{1.25}{1.14} \pm 0.12$ & $16.56 \aerr{1.79}{1.70} \pm 0.20$ & $4.26 \aerr{0.21}{0.20} \pm 0.04$ & $3.58 \aerr{0.20}{0.19} \pm 0.04$ & $3.66 \pm 0.09\pm0.03$ \\
     2  & $\phantom{1}2.41 \aerr{1.11}{0.98} \pm 0.14$ & $24.12 \aerr{2.30}{2.19} \pm 0.28$ & $4.52 \aerr{0.24}{0.23} \pm 0.05$ & $3.68 \aerr{0.22}{0.22} \pm 0.04$ & $4.32 \pm 0.10\pm0.04$ \\  
     3  &           $11.99 \aerr{1.76}{1.64} \pm 0.19$ & $20.91 \aerr{2.12}{2.01} \pm 0.25$ & $4.29 \aerr{0.23}{0.22} \pm 0.05$ & $3.59 \aerr{0.22}{0.21} \pm 0.04$ & $3.73 \pm 0.10\pm0.04$\\
     4  &           $13.71 \aerr{1.55}{1.47} \pm 0.17$ & $10.84 \aerr{1.45}{1.36} \pm 0.16$ & $1.94 \aerr{0.15}{0.14} \pm 0.03$ & $2.11 \aerr{0.15}{0.15} \pm 0.03$ & $2.25 \pm 0.07\pm0.02$ \\
     \bottomrule
  \end{tabular}
  }
\end{table} 
\FloatBarrier 
\section{Interpretation and conclusions} 

\label{sec:interpretationFit} 
The observables presented in the previous section are interpreted in terms of $\gamma$ and related hadronic parameters by minimising the $\chi^2$ of the observables with respect to the physics parameters. The $\chi^2$ is
\begin{equation}
  \chi^2_{B\to Dh} = \sum_{ij} [\mathbf{V}^{-1}]_{ij} \left( x_i - \hat{x}_i \right) \left( x_j - \hat{x}_j \right),
\end{equation}
where $x_i$ is the measured value of the $i$th observable and $\hat{x}_i$ is the corresponding expectation value, which in turn depends on the physics parameters. 
The expected values are corrected for the small difference in efficiencies between the OS and LS decay modes using large samples of simulated decays, with the largest deviation from unity for such factors being around $1\%$. The covariance matrix, $\mathbf{V}$, is given by the sum of the statistical and systematic covariance matrices, where the statistical covariance matrix is given by 
\begin{equation}
  [\mathbf{V}_{\rm{stat}.}]_{ij} = \rho_{ij} \sigma_i( x_i - \hat{x}_i ) \sigma_j(x_j - \hat{x}_j) ,
\end{equation}
and $\rho_{ij}$ is the statistical correlation between observables $x_i$ and $x_j$. The uncertainties, $\sigma_i$, are parameterised as prescribed in Ref.~\cite{Barlow:2004wg} by assuming the variance is a linear function, which gives
\begin{equation}
  \sigma( x - \hat{x} ) = \sqrt{ \sigma_- \sigma_+  + (\sigma_+-\sigma_-) ( x - \hat{x} ) },
\end{equation}
where $\sigma_{\pm}$ are the positive and negative uncertainties on each parameter given by the fit described in Sect.~\ref{sec:massFit}. 
The robustness of the determination of the physics parameters is verified by repeating the fit on a large ensemble of simulated samples. 
The assigned uncertainties and central values are found to describe the ensemble very well. 
However, a bias is found on $r_B^\pi$ of around 10\% of the its uncertainty, and is a consequence of the behaviour of the observables as $r_B^\pi \rightarrow 0$. 
Thus the parameterisation 
\begin{equation}
  x_{Dh} + i y_{Dh}  = r_B^h e^{ i \delta_B^h }
\end{equation}
is adopted to facilitate the combination of these results with other decay modes.
The Cartesian form of the $\Bpm \rightarrow D \Kpm$ decay parameters is also reported for consistency.
The charm mixing parameters are constrained using the values reported in Ref.~\cite{Amhis:2019ckw}.
The hadronic parameters of the $D$ decay are constrained using a combination of results from BESIII, CLEO-c and LHCb experiments~\cite{LHCb-PAPER-2015-057,Evans:2019wza,BESIII:2021eud}. 
The full BESIII $\chi^2$ and CLEO-c likelihoods are used in the constraint, owing to the non-Gaussian nature of the uncertainties on these parameters, and are calculated using the supplementary material provided in Ref.~\cite{BESIII:2021eud}. 
The fit is repeated fixing the $D$ hadronic parameters to the values predicted by the models presented in Ref.~\cite{LHCb-PAPER-2017-040} in order to assess the contribution to the uncertainties from the limited knowledge on these parameters.
The $B$-hadronic parameters, in the polar form, are found to be 
\begin{equation*}
  \begin{split}
    \delta_B^K   &=  \dBKv, \\
    \delta_B^\pi &=  \dBpv, \\
    r_B^K        &=  \rBKv, \\
    r_B^\pi      &=  \rBpv, 
  \end{split}
\end{equation*}
where the first and second uncertainties are statistical and systematic, respectively, while the third is from the finite knowledge on the $D$-meson decay parameters. 
The hadronic parameters in the Cartesian form are found to be 
\begin{equation*}
  \begin{split}
    x_{DK}   &= \xkv, \\ 
    y_{DK}   &= \ykv, \\ 
    x_{D\pi} &= \xpiv, \\
    y_{D\pi} &= \ypiv. \\ 
  \end{split}
  \end{equation*}
The value of $\gamma$ is found to be 
\begin{equation*}
  \gamma = \gammav, 
\end{equation*}
which is one of the most precise determinations thus far using any single $D$-decay mode, and is compatible with current averages \cite{Amhis:2019ckw, LHCb:2021dcr}. The correlation matrix of the fit to $\gamma$ and the $B$-hadronic parameters is given in Table~\ref{tb:correlation}.

\begin{table}
    \centering
        \caption{\label{tb:correlation} Correlation matrix of $\gamma$ with the $B$-hadronic parameters, expressed in the Cartesian form. }
\begin{tabular}{l | rrrrr}
\toprule
 & $\gamma$ & $x_{DK}$ & $y_{DK}$ & $x_{D\pi}$ & $y_{D\pi}$\\
\midrule
$\gamma$ & $1\phantom{.000}$ & $-0.286$ & $-0.305$ & $-0.196$ & $0.047$\\
$x_{DK}$ &  & $1\phantom{.000}$ & $0.911$ & $-0.118$ & $-0.211$\\
$y_{DK}$ &  &  & $1\phantom{.000}$ & $-0.076$ & $-0.173$\\
$x_{D\pi}$ &  &  &  & $1\phantom{.000}$ & $0.357$\\
$y_{D\pi}$ &  &  &  &  & $1\phantom{.000}$\\
\bottomrule
\end{tabular}
\end{table}

In summary, the analysis presented in this paper gives the first measurement of parameters of interest in $\Bpm \to D \left[ \Kmp \pipm\pipm\pimp \right] h^\pm$ decays in bins of the phase space of the $D$ decay, where the magnitude of the $\CP$ violation in one of the bins is the largest yet observed. Furthermore, the obtained results are anticipated to have a strong impact on the overall knowledge of $\gamma$.

\section*{Acknowledgements}
%
%
\noindent We express our gratitude to our colleagues in the CERN
accelerator departments for the excellent performance of the LHC. We
thank the technical and administrative staff at the LHCb
institutes.
We acknowledge support from CERN and from the national agencies:
CAPES, CNPq, FAPERJ and FINEP (Brazil); 
MOST and NSFC (China); 
CNRS/IN2P3 (France); 
BMBF, DFG and MPG (Germany); 
INFN (Italy); 
NWO (Netherlands); 
MNiSW and NCN (Poland); 
MEN/IFA (Romania); 
MICINN (Spain); 
SNSF and SER (Switzerland); 
NASU (Ukraine); 
STFC (United Kingdom); 
DOE NP and NSF (USA).
We acknowledge the computing resources that are provided by CERN, IN2P3
(France), KIT and DESY (Germany), INFN (Italy), SURF (Netherlands),
PIC (Spain), GridPP (United Kingdom), 
CSCS (Switzerland), IFIN-HH (Romania), CBPF (Brazil),
Polish WLCG  (Poland) and NERSC (USA).
We are indebted to the communities behind the multiple open-source
software packages on which we depend.
Individual groups or members have received support from
ARC and ARDC (Australia);
Minciencias (Colombia);
AvH Foundation (Germany);
EPLANET, Marie Sk\l{}odowska-Curie Actions and ERC (European Union);
A*MIDEX, ANR, IPhU and Labex P2IO, and R\'{e}gion Auvergne-Rh\^{o}ne-Alpes (France);
Key Research Program of Frontier Sciences of CAS, CAS PIFI, CAS CCEPP, 
Fundamental Research Funds for the Central Universities, 
and Sci. \& Tech. Program of Guangzhou (China);
GVA, XuntaGal, GENCAT and Prog.~Atracci\'on Talento, CM (Spain);
SRC (Sweden);
the Leverhulme Trust, the Royal Society
 and UKRI (United Kingdom).

\appendix
\renewcommand{\thesection}{\Alph{section}}

\section{Analysis of $\boldsymbol{X_b \to D^{*+} \mu^- \neumb X}$}
\label{app:DT}

Flavour-tagged $\Dz$ mesons are provided by a sample of ${X_b \to D^{*+} \mu^- \neumb X}$ decays. 
The configuration where the kaon from the $D$ decay are of the same charge to the pion from the $D^{*+}$ decay is referred to as \textit{wrong sign} (WS). 
WS decays mostly proceed via the DCS amplitude, with a small additional contribution from charm mixing. 
The configuration where the kaon is of the same sign as the pion from the $D^{*+}$ decay are dominated by CF transitions, and provide the normalisation channel for the $\mathcal{R}^{i}_{DT}$ observables. 
Inclusion of such observables gives significant reduction in the uncertainties on the $r^i_{K3\pi}$ parameters, and as a consequence a more precise determination on $\delta_B^\pi$.

The selection of the doubly tagged data set is simpler than the fully hadronic mode, as a pure sample can be selected using only requirements on PID variables of the $D$-decay products and on the $D$-candidate mass. 
The requirements used in the selection of $\Bpm\to D h^\pm$ decays to remove both crossfeed from the CF mode and candidates where one of the tracks has been duplicated are also applied.

\begin{figure}
    \centering
    \includegraphics[width=0.95\textwidth]{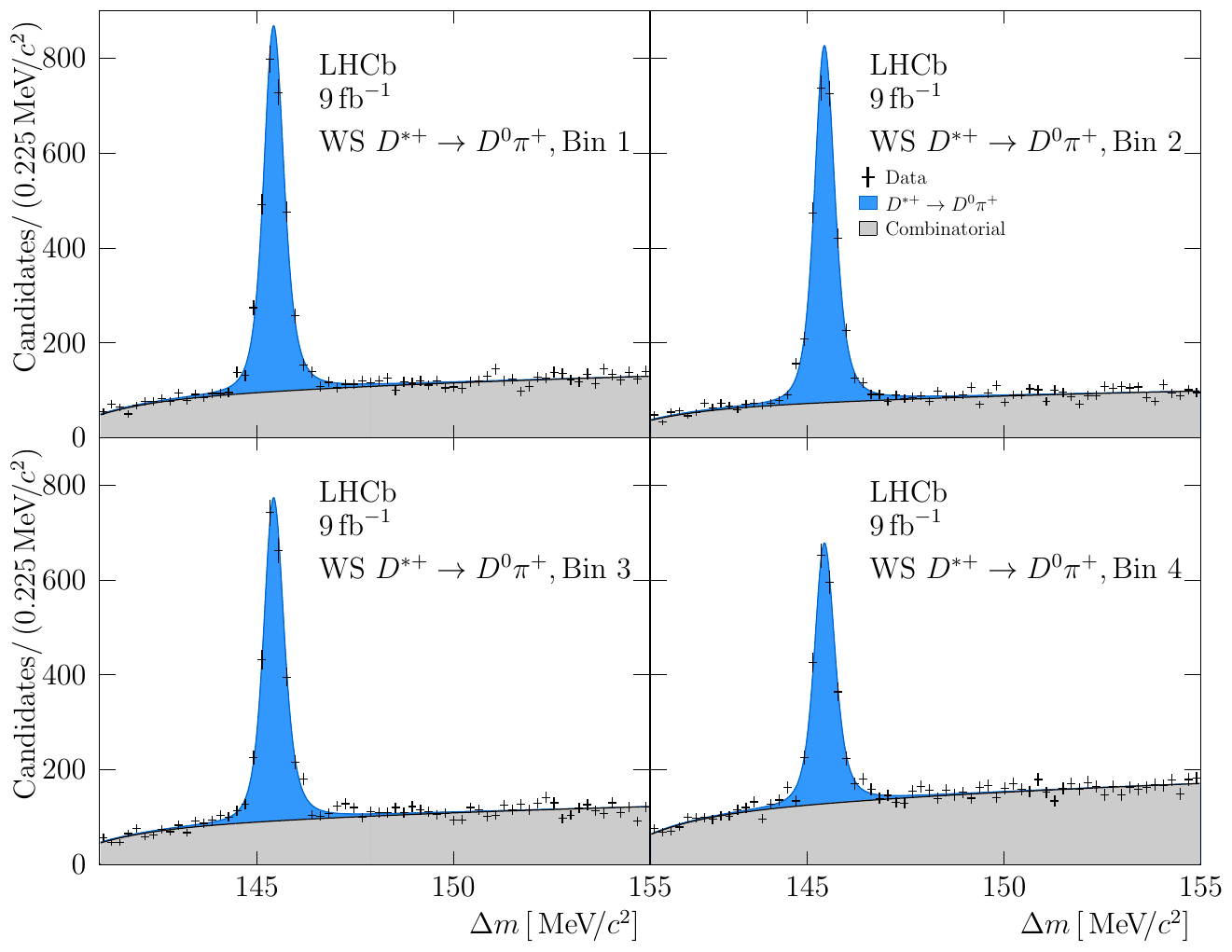}
    \caption{$\Delta m$ distributions for WS $D^{*+} \to \Dz \pip$ candidates,
    where the $D^{*+}$ has been produced via ${X_b \to D^{*+} \mu^- \neumb X}$ decays. The candidates are divided by $D$-decay phase-space bin.}
    \label{fig:sl_fit}
\end{figure}

Signal yields, and their corresponding uncertainties, are obtained by performing a fit to the mass difference between the $D^{*+}$ and $\Dz$ candidates, referred to as $\Delta m$. 
The signal is described by the same parameterisation as the fully reconstructed $\Bpm \to D h^\pm$ decays, while the only significant background is combinatorial, and can be described by a threshold function
\begin{equation}
 \mathcal{P}_\text{bkg}( \Delta m ) \propto \left( 1 - e^{c( \Delta m - m_\pi ) } \right) \left( \Delta m - m_\pi \right)^{a},
\end{equation}
where $a$ and $c$ are parameters that vary freely in the fit, and $m_\pi$ is the mass of the pion. The distribution of $\Delta m$ for ${X_b \to D^{*+} \mu^- \neumb X}$ decays, in the four bins of phase space, is shown in Fig.~\ref{fig:sl_fit}.
The first three bins have a similar population of both signal and background, while the fourth bin has a smaller signal yield due to the smaller DCS amplitude in this bin. 
The background level is also slightly larger in this bin, due to both the larger CF amplitude and phase-space volume associated with this region. 

\FloatBarrier

\section{Additional material used in interpretation} 

\begin{table}
\caption{\label{stat_cov_RK}Statistical correlation matrix between $\mathcal{R}_{h^\pm}^i$ observables}
\centering
\scalebox{0.60}{
\def\arraystretch{1.2}
\begin{tabular}{l|rrrrrrrrrrrrrrrr}
\toprule
 $\rho_\text{stat.}$ & $\mathcal{R}_{\Kp}^1$ & $\mathcal{R}_{\Km}^1$ & $\mathcal{R}_{\pip}^1$ & $\mathcal{R}_{\pim}^1$ & $\mathcal{R}_{\Kp}^2$ & $\mathcal{R}_{\Km}^2$ & $\mathcal{R}_{\pip}^2$ & $\mathcal{R}_{\pim}^2$ & $\mathcal{R}_{\Kp}^3$ & $\mathcal{R}_{\Km}^3$ & $\mathcal{R}_{\pip}^3$ & $\mathcal{R}_{\pim}^3$ & $\mathcal{R}_{\Kp}^4$ & $\mathcal{R}_{\Km}^4$ & $\mathcal{R}_{\pip}^4$ & $\mathcal{R}_{\pim}^4$\\
\midrule
$\mathcal{R}_{\Kp}^1$ & $1.000$ & $0.004$ & $-0.072$ & $0.002$ & $-0.036$ & $0.001$ & $0.003$ & $0.000$ & $-0.001$ & $0.001$ & $0.000$ & $0.000$ & $-0.009$ & $0.001$ & $0.001$ & $0.000$\\
$\mathcal{R}_{\Km}^1$ &  & $1.000$ & $0.002$ & $-0.059$ & $0.001$ & $-0.036$ & $0.000$ & $0.002$ & $0.001$ & $-0.001$ & $0.000$ & $0.000$ & $0.001$ & $-0.009$ & $0.000$ & $0.001$\\
$\mathcal{R}_{\pip}^1$ &  &  & $1.000$ & $0.005$ & $0.003$ & $0.000$ & $-0.035$ & $0.000$ & $0.000$ & $0.000$ & $-0.000$ & $0.001$ & $0.001$ & $0.000$ & $-0.009$ & $0.001$\\
$\mathcal{R}_{\pim}^1$ &  &  &  & $1.000$ & $0.000$ & $0.002$ & $0.000$ & $-0.035$ & $0.000$ & $0.000$ & $0.001$ & $-0.000$ & $0.000$ & $0.001$ & $0.001$ & $-0.010$\\
$\mathcal{R}_{\Kp}^2$ &  &  &  &  & $1.000$ & $0.004$ & $-0.087$ & $0.002$ & $-0.041$ & $0.001$ & $0.004$ & $0.000$ & $-0.001$ & $0.001$ & $0.000$ & $0.000$\\
$\mathcal{R}_{\Km}^2$ &  &  &  &  &  & $1.000$ & $0.001$ & $-0.040$ & $0.001$ & $-0.045$ & $-0.000$ & $0.003$ & $0.001$ & $0.000$ & $0.000$ & $0.000$\\
$\mathcal{R}_{\pip}^2$ &  &  &  &  &  &  & $1.000$ & $0.005$ & $0.004$ & $0.000$ & $-0.040$ & $0.000$ & $0.000$ & $0.000$ & $-0.001$ & $0.001$\\
$\mathcal{R}_{\pim}^2$ &  &  &  &  &  &  &  & $1.000$ & $0.000$ & $0.002$ & $0.000$ & $-0.040$ & $0.000$ & $0.000$ & $0.001$ & $-0.001$\\
$\mathcal{R}_{\Kp}^3$ &  &  &  &  &  &  &  &  & $1.000$ & $0.004$ & $-0.080$ & $0.002$ & $-0.036$ & $0.001$ & $0.003$ & $0.000$\\
$\mathcal{R}_{\Km}^3$ &  &  &  &  &  &  &  &  &  & $1.000$ & $0.002$ & $-0.066$ & $0.001$ & $-0.035$ & $0.000$ & $0.003$\\
$\mathcal{R}_{\pip}^3$ &  &  &  &  &  &  &  &  &  &  & $1.000$ & $0.005$ & $0.003$ & $0.000$ & $-0.036$ & $0.001$\\
$\mathcal{R}_{\pim}^3$ &  &  &  &  &  &  &  &  &  &  &  & $1.000$ & $0.000$ & $0.002$ & $0.001$ & $-0.037$\\
$\mathcal{R}_{\Kp}^4$ &  &  &  &  &  &  &  &  &  &  &  &  & $1.000$ & $0.005$ & $-0.071$ & $0.002$\\
$\mathcal{R}_{\Km}^4$ &  &  &  &  &  &  &  &  &  &  &  &  &  & $1.000$ & $0.002$ & $-0.069$\\
$\mathcal{R}_{\pip}^4$ &  &  &  &  &  &  &  &  &  &  &  &  &  &  & $1.000$ & $0.006$\\
$\mathcal{R}_{\pim}^4$ &  &  &  &  &  &  &  &  &  &  &  &  &  &  &  & $1.000$\\
\bottomrule
\end{tabular}
}
\end{table}

\begin{table}
\caption{\label{sys_cov_RK}Systematic correlation matrix between $\mathcal{R}_{h^\pm}^i$ observables}
\centering
\scalebox{0.60}{
\def\arraystretch{1.2}
\begin{tabular}{l|rrrrrrrrrrrrrrrr}
\toprule
 $\rho_\text{sys.}$  & $\mathcal{R}_{\Kp}^1$ & $\mathcal{R}_{\Km}^1$ & $\mathcal{R}_{\pip}^1$ & $\mathcal{R}_{\pim}^1$ & $\mathcal{R}_{\Kp}^2$ & $\mathcal{R}_{\Km}^2$ & $\mathcal{R}_{\pip}^2$ & $\mathcal{R}_{\pim}^2$ & $\mathcal{R}_{\Kp}^3$ & $\mathcal{R}_{\Km}^3$ & $\mathcal{R}_{\pip}^3$ & $\mathcal{R}_{\pim}^3$ & $\mathcal{R}_{\Kp}^4$ & $\mathcal{R}_{\Km}^4$ & $\mathcal{R}_{\pip}^4$ & $\mathcal{R}_{\pim}^4$\\
\midrule
$\mathcal{R}_{\Kp}^1$ & $1.000$ & $-0.144$ & $0.569$ & $0.034$ & $0.423$ & $0.030$ & $0.300$ & $-0.225$ & $0.340$ & $0.015$ & $0.275$ & $-0.175$ & $0.195$ & $0.021$ & $0.220$ & $-0.146$\\
$\mathcal{R}_{\Km}^1$ &  & $1.000$ & $0.004$ & $0.353$ & $-0.185$ & $0.160$ & $-0.157$ & $0.206$ & $-0.131$ & $0.236$ & $-0.136$ & $0.170$ & $0.023$ & $0.180$ & $-0.096$ & $0.148$\\
$\mathcal{R}_{\pip}^1$ &  &  & $1.000$ & $0.297$ & $0.294$ & $-0.036$ & $0.367$ & $-0.325$ & $0.293$ & $-0.157$ & $0.359$ & $-0.269$ & $0.234$ & $-0.202$ & $0.332$ & $-0.177$\\
$\mathcal{R}_{\pim}^1$ &  &  &  & $1.000$ & $-0.244$ & $0.093$ & $-0.319$ & $0.342$ & $-0.224$ & $0.212$ & $-0.299$ & $0.311$ & $-0.166$ & $0.263$ & $-0.254$ & $0.259$\\
$\mathcal{R}_{\Kp}^2$ &  &  &  &  & $1.000$ & $-0.168$ & $0.579$ & $0.056$ & $0.405$ & $-0.101$ & $0.249$ & $-0.212$ & $0.260$ & $0.021$ & $0.221$ & $-0.130$\\
$\mathcal{R}_{\Km}^2$ &  &  &  &  &  & $1.000$ & $0.039$ & $0.146$ & $-0.053$ & $0.173$ & $-0.052$ & $0.071$ & $0.036$ & $0.142$ & $-0.034$ & $0.056$\\
$\mathcal{R}_{\pip}^2$ &  &  &  &  &  &  & $1.000$ & $0.335$ & $0.267$ & $-0.162$ & $0.306$ & $-0.302$ & $0.232$ & $-0.159$ & $0.340$ & $-0.154$\\
$\mathcal{R}_{\pim}^2$ &  &  &  &  &  &  &  & $1.000$ & $-0.272$ & $0.211$ & $-0.344$ & $0.251$ & $-0.182$ & $0.264$ & $-0.232$ & $0.264$\\
$\mathcal{R}_{\Kp}^3$ &  &  &  &  &  &  &  &  & $1.000$ & $-0.147$ & $0.655$ & $0.231$ & $0.297$ & $-0.044$ & $0.223$ & $-0.143$\\
$\mathcal{R}_{\Km}^3$ &  &  &  &  &  &  &  &  &  & $1.000$ & $-0.023$ & $0.311$ & $-0.076$ & $0.306$ & $-0.139$ & $0.158$\\
$\mathcal{R}_{\pip}^3$ &  &  &  &  &  &  &  &  &  &  & $1.000$ & $0.439$ & $0.207$ & $-0.194$ & $0.302$ & $-0.176$\\
$\mathcal{R}_{\pim}^3$ &  &  &  &  &  &  &  &  &  &  &  & $1.000$ & $-0.172$ & $0.240$ & $-0.220$ & $0.215$\\
$\mathcal{R}_{\Kp}^4$ &  &  &  &  &  &  &  &  &  &  &  &  & $1.000$ & $-0.001$ & $0.613$ & $0.385$\\
$\mathcal{R}_{\Km}^4$ &  &  &  &  &  &  &  &  &  &  &  &  &  & $1.000$ & $-0.033$ & $0.282$\\
$\mathcal{R}_{\pip}^4$ &  &  &  &  &  &  &  &  &  &  &  &  &  &  & $1.000$ & $0.593$\\
$\mathcal{R}_{\pim}^4$ &  &  &  &  &  &  &  &  &  &  &  &  &  &  &  & $1.000$\\
\bottomrule
\end{tabular}
}
\end{table}

\label{app:covAndEfficiency}
This section gives additional material used in the interpretation of $\mathcal{R}_{h^\pm}^i$ and $\mathcal{R}_\text{DT}^i$ in terms of $\gamma$ and the hadronic parameters. The statistical and systematic correlation matrices for $\mathcal{R}^i_{h^{\pm}}$ are given in Tables~\ref{stat_cov_RK} and \ref{sys_cov_RK}, respectively. The observables are found to be  largely statistically uncorrelated, while the systematic correlations are considerable. The corresponding matrices for the double-tag observables are given in Tables~\ref{stat_cov_DT} and ~\ref{sys_cov_DT}. 

The corrections to the observables, denoted by $\kappa$, due to the variation in efficiencies across the phase space are given in Table~\ref{tb:effCorr}.
The quoted uncertainties are due to the finite size of the simulated samples used to evaluate the corrections.
The derivatives of the corrections with respect to the parameters of interest are also given such that this dependence can be included in the interpretation of the results. 

\begin{table}
\caption{\label{stat_cov_DT}Statistical correlation matrix between $\mathcal{R}_\text{DT}^i$ observables}
\centering
\def\arraystretch{1.2}

\begin{tabular}{r
r
r
r
r
}
\toprule
 & $\mathcal{R}_{\text{DT}}^1$ & $\mathcal{R}_{\text{DT}}^2$ & $\mathcal{R}_{\text{DT}}^3$ & $\mathcal{R}_{\text{DT}}^4$\\
\midrule
$\mathcal{R}_{\text{DT}}^1$ & $1.000$ & $-0.018$ & $0.019$ & $0.017$\\
$\mathcal{R}_{\text{DT}}^2$ &  & $1.000$ & $-0.024$ & $0.020$\\
$\mathcal{R}_{\text{DT}}^3$ &  &  & $1.000$ & $-0.009$\\
$\mathcal{R}_{\text{DT}}^4$ &  &  &  & $1.000$\\
\bottomrule
\end{tabular}
\end{table}

\begin{table}
\caption{\label{sys_cov_DT}Systematic correlation matrix between $\mathcal{R}_\text{DT}^i$ observables}
\centering
\def\arraystretch{1.2}
\begin{tabular}{r
r
r
r
r
}
\toprule
 & $\mathcal{R}_{\text{DT}}^1$ & $\mathcal{R}_{\text{DT}}^2$ & $\mathcal{R}_{\text{DT}}^3$ & $\mathcal{R}_{\text{DT}}^4$\\
\midrule
$\mathcal{R}_{\text{DT}}^1$ & $1.000$ & $-0.149$ & $-0.100$ & $0.018$\\
$\mathcal{R}_{\text{DT}}^2$ &  & $1.000$ & $-0.023$ & $-0.026$\\
$\mathcal{R}_{\text{DT}}^3$ &  &  & $1.000$ & $-0.011$\\
$\mathcal{R}_{\text{DT}}^4$ &  &  &  & $1.000$\\
\bottomrule
\end{tabular}
\end{table}

\begin{table}
\caption{\label{tb:effCorr}Efficiency corrections to the observables, where the quoted uncertainties are due to the finite sizes of the simulated samples used to evaluate the corrections. The derivatives of the corrections with respect to the $\gamma$ and the $B$-decay parameters, using $z_{h^\pm} \equiv r_B^h e^{i (\delta_B^h \pm \gamma)}$, and the charm mixing parameters $(x,y)$ are also given. }
\centering
\def\arraystretch{1.2}

\begin{tabular}{l
>{\collectcell\num}r<{\endcollectcell} @{${}\pm{}$} >{\collectcell\num}l<{\endcollectcell}
cccc}
\toprule

                      & \multicolumn{2}{c}{$\kappa_0$} 
                      &  \multicolumn{1}{c}{$\displaystyle \frac{ \partial\kappa}{ \partial \text{Re}( z_{h^\pm} ) } $} 
                      &  \multicolumn{1}{c}{$\displaystyle \frac{ \partial\kappa}{ \partial \text{Im}( z_{h^\pm} ) } $} 
                      &  \multicolumn{1}{c}{$\displaystyle \frac{ \partial\kappa}{ \partial x } $} 
                                            & \multicolumn{1}{c}{$\displaystyle \frac{ \partial\kappa}{ \partial y } $} \\
                      \midrule
$\kappa_{\Kp}^1$      & 0.992 & 0.002 &  $-0.042$  & $-0.044$             & $-0.686$  & $-0.993$  \\
$\kappa_{\Kp}^2$      & 0.991 & 0.002 &  $-0.055$  & $-0.015$             & $-0.372$  & $-0.980$  \\
$\kappa_{\Kp}^3$      & 0.991 & 0.002 &  $-0.065$  & $\phantom{-}0.012 $  & $-0.132$  & $-1.103$  \\
$\kappa_{\Kp}^4$      & 0.990 & 0.002 &  $-0.073$  & $\phantom{-}0.090 $  & $-0.071$  & $-1.475$  \\
\midrule
$\kappa_{\Km}^1$      & 1.008 & 0.003 &  $\phantom{-}0.120 $  & $-0.058$  & $\phantom{-}1.771 $  & $\phantom{-}1.235$ \\
$\kappa_{\Km}^2$      & 1.006 & 0.003 &  $\phantom{-}0.305 $  & $-0.023$  & $\phantom{-}2.315 $  & $-0.360$  \\
$\kappa_{\Km}^3$      & 1.004 & 0.003 &  $\phantom{-}0.131 $  & $-0.024$  & $\phantom{-}1.654 $  & $-0.093$  \\
$\kappa_{\Km}^4$      & 1.011 & 0.003 &  $\phantom{-}0.027 $  & $-0.094$  & $\phantom{-}1.404 $  & $\phantom{-}0.457$   \\
\midrule
$\kappa_{\pip}^1$     & 0.989 & 0.003 &  $\phantom{-}0.223 $  & $-0.137$  & $-1.855$  & $-0.553$  \\
$\kappa_{\pip}^2$     & 0.981 & 0.003 &  $-0.147$             & $-0.047$  & $-0.812$  & $-1.981$  \\
$\kappa_{\pip}^3$     & 0.985 & 0.003 &  $-0.013$             & $\phantom{-}0.289 $  & $\phantom{-}0.697 $  & $-2.024$  \\
$\kappa_{\pip}^4$     & 1.013 & 0.003 &  $\phantom{-}0.558 $  & $\phantom{-}0.102 $  & $\phantom{-}2.210 $  & $-0.431$  \\
\midrule
$\kappa_{\pim}^1$     & 0.988 & 0.003 &  $\phantom{-}0.129 $  & $-0.160$  & $-1.725$  & $-0.844$  \\
$\kappa_{\pim}^2$     & 0.982 & 0.003 &  $-0.150$             & $-0.049$  & $-0.747$  & $-1.835$  \\
$\kappa_{\pim}^3$     & 0.985 & 0.003 &  $-0.084$             & $\phantom{-}0.215 $  & $\phantom{-}0.444 $  & $-1.973$  \\
$\kappa_{\pim}^4$     & 1.007 & 0.003 &  $\phantom{-}0.598 $  & $\phantom{-}0.271 $  & $\phantom{-}1.882 $  & $-1.123$  \\
\midrule
$\kappa_\text{DT}^1$  & 1.008 & 0.003 &          &         & $-0.054$  & $\phantom{-}0.083$   \\
$\kappa_\text{DT}^2$  & 1.002 & 0.003 &          &         & $\phantom{-}0.047$   & $\phantom{-}0.083$   \\
$\kappa_\text{DT}^3$  & 1.001 & 0.003 &          &         & $\phantom{-}0.119$   & $-0.006$  \\
$\kappa_\text{DT}^4$  & 1.001 & 0.003 &          &         & $\phantom{-}0.120$   & $\phantom{-}0.191$   \\

\bottomrule
\end{tabular}
\end{table}

\FloatBarrier

\addcontentsline{toc}{section}{References}
\bibliographystyle{LHCb}
\bibliography{main,standard,LHCb-PAPER,LHCb-CONF,LHCb-DP,LHCb-TDR}
\newpage
\centerline
{\large\bf LHCb collaboration}
\begin
{flushleft}
\small
R.~Aaij$^{32}$\lhcborcid{0000-0003-0533-1952},
A.S.W.~Abdelmotteleb$^{50}$\lhcborcid{0000-0001-7905-0542},
C.~Abellan~Beteta$^{44}$,
F.~Abudin{\'e}n$^{50}$\lhcborcid{0000-0002-6737-3528},
T.~Ackernley$^{54}$\lhcborcid{0000-0002-5951-3498},
B.~Adeva$^{40}$\lhcborcid{0000-0001-9756-3712},
M.~Adinolfi$^{48}$\lhcborcid{0000-0002-1326-1264},
H.~Afsharnia$^{9}$,
C.~Agapopoulou$^{13}$\lhcborcid{0000-0002-2368-0147},
C.A.~Aidala$^{77}$\lhcborcid{0000-0001-9540-4988}
% ...

{\footnotesize \it

% ...
$^{9}$Universit{\'e} Clermont Auvergne, CNRS/IN2P3, LPC, Clermont-Ferrand, France\\
% ...
$^{13}$LPNHE, Sorbonne Universit{\'e}, Paris Diderot Sorbonne Paris Cit{\'e}, CNRS/IN2P3, Paris, France\\
% ...
$^{32}$Nikhef National Institute for Subatomic Physics, Amsterdam, Netherlands\\
% ...
$^{40}$Instituto Galego de F{\'\i}sica de Altas Enerx{\'\i}as (IGFAE), Universidade de Santiago de Compostela, Santiago de Compostela, Spain\\
% ...
$^{44}$Physik-Institut, Universit{\"a}t Z{\"u}rich, Z{\"u}rich, Switzerland\\
% ...
$^{48}$H.H. Wills Physics Laboratory, University of Bristol, Bristol, United Kingdom\\
% ...
$^{50}$Department of Physics, University of Warwick, Coventry, United Kingdom\\
% ...
$^{54}$Oliver Lodge Laboratory, University of Liverpool, Liverpool, United Kingdom\\
% ...
$^{62}$Syracuse University, Syracuse, NY, United States\\
% ...
$^{77}$University of Michigan, Ann Arbor, MI, United States, associated to $^{62}$\\
\bigskip
% ...
}
\end{flushleft}




\end{document}